\documentclass[conference]{IEEEtran}

\usepackage{amsmath}
\usepackage{amssymb}  
\usepackage{graphics}    
\usepackage{graphicx}
\usepackage{color}
\usepackage{pgfplots}
\usepackage{tikz}
\DeclareMathAlphabet{\mathcalligra}{T1}{calligra}{m}{n}
\usetikzlibrary{arrows,petri,topaths,backgrounds,snakes,patterns,positioning}
\usepackage{hyperref}
\usepackage{setspace}
\usepackage{tkz-berge}
\usetikzlibrary{pgfplots.groupplots}
\ifCLASSOPTIONcompsoc \usepackage[caption=false,font=normalsize,labelfont=sf,textfont=sf]{subfig} \else \usepackage[caption=false,font=footnotesize]{subfig}\fi
\usepackage{xcolor}
\usepackage{float}
\usepackage{algorithmic}
\usepackage[linesnumbered,lined,boxed]{algorithm2e}
\usepackage{tabularx} 
\usepackage{booktabs}
\usepackage{multirow}
\usepackage{mathtools}
\usepackage{setspace}
\usepackage{placeins}
\newcommand{\ApplyGradient}[1]{%
	\pgfmathsetmacro{\PercentColor}{100.0*(#1-\MinNumber)/(\MaxNumber-\MinNumber)}
	\hspace{-0.33em}\colorbox{red!\PercentColor!yellow}{}
}

\newcolumntype{R}{>{\collectcell\ApplyGradient}c<{\endcollectcell}}
\renewcommand{\arraystretch}{0}
\usepackage{collcell}
\pgfplotsset{compat=1.6, ylabsh/.style={every axis y label/.style={at={(0,0.5)}}}}

\usepackage{adjustbox}
\usetikzlibrary{fit}
\tikzset{%
	highlight/.style={rectangle,rounded corners,fill=red!15,draw,fill opacity=0.5,thick,inner sep=0pt}
}

\newcommand{\dst}{\displaystyle}

\newcommand*{\MinNumber}{0}%
\newcommand*{\MaxNumber}{1}%
\makeatletter\def\@IEEEpubidpullup{8\baselineskip}\makeatother
\begin{document}

\title{GEMSEC: Graph Embedding with Self Clustering}
\author{\IEEEauthorblockN{Benedek Rozemberczki, Ryan Davies, Rik Sarkar and Charles Sutton}
\IEEEauthorblockA{\textit{School of Informatics, The University of Edinburgh} \\
\{benedek.rozemberczki,ryan.davies\}@ed.ac.uk,\{rsarkar,csutton\}@inf.ed.ac.uk}}

%E-mail: see http://www.michaelshell.org/contact.html
%\IEEEcompsocthanksitem J. Doe and J. Doe are with Anonymous University.}

\IEEEtitleabstractindextext{%
\begin{abstract}
Modern graph embedding procedures can efficiently process graphs with millions of nodes. In this paper, we propose \textit{GEMSEC} -- a graph embedding algorithm which learns a clustering of the nodes simultaneously with computing their embedding. 
\textit{GEMSEC} is a general extension of earlier work in the domain of sequence-based graph embedding. \textit{GEMSEC} places nodes in an abstract feature space where the vertex features minimize the negative log-likelihood of preserving sampled vertex neighborhoods, and it incorporates known social network properties through a machine learning regularization. We present two new social network datasets and show that by simultaneously considering the embedding and clustering problems with respect to social properties, \textit{GEMSEC} extracts high-quality clusters competitive with or superior to other community detection algorithms. In experiments, the method is found to be computationally efficient and robust to the choice of hyperparameters. 
\end{abstract}
}

\maketitle

\IEEEdisplaynontitleabstractindextext

\IEEEpeerreviewmaketitle

\section{Introduction}\label{sec:introduction}
Community detection is one of the most important problems in network analysis due to its wide applications ranging from the analysis of collaboration networks to image segmentation, the study of protein-protein interaction networks in biology, and many others~\cite{van2012robust,backstrom2006group,papadopoulos2012community}. 
Communities are usually defined as groups of nodes that are connected to each other more densely than to the rest of the network. Classical approaches to community detection depend on properties such as graph metrics, spectral properties and density of shortest paths~\cite{leskovec2014mining}. Random walks and randomized label propagation~\cite{walktrap,gregory2010finding} have also been investigated.

Embedding the nodes in a low dimensional Euclidean space enables us to apply standard machine learning techniques. This space is sometimes called the {\em feature space} -- implying that it represents abstract structural features of the network. %For the transportation network, the geographic distance between nodes may be preserved. 
Embeddings have been used for machine learning tasks such as labeling nodes, regression, link prediction, and graph visualization, see~\cite{goyal2017graph} for a survey. Graph embedding processes usually aim to preserve certain predefined differences between nodes encoded in their embedding distances. 
For social network embedding, a natural priority is to preserve community membership and enable community detection. 

Recently, sequence-based methods have been developed as a way to convert complex, non-linear network structures into formats more compatible with vector spaces. These methods sample sequences of nodes from the graph using a randomized mechanism (e.g. random walks), with the idea that nodes that are ``close'' in the graph connectivity will also frequently appear close in a sampling of random walks. The methods then proceed to use this random-walk-proximity information as a basis to embed nodes such that socially close nodes are placed nearby. In this category, \textit{Deepwalk}~\cite{deepwalk} and \textit{Node2Vec}~\cite{grover2016node2vec} are two popular methods.

\begin{figure}[h!]
	\centering
	\subfloat[DeepWalk]{\begin{tikzpicture}[scale=0.3,transform shape]
		\tikzstyle{VertexStyle}=[minimum size = 15pt,inner sep=0pt, shape = circle]
		\Vertex[L=$$,x=-5.09505548965,y=-9.45204196694]{0}
		\Vertex[L=$$,x=-3.01140568947,y=-6.64526573061]{1}
		\Vertex[L=$$,x=-6.23543840057,y=-5.34444315124]{2}
		\Vertex[L=$$,x=-10.0279450316,y=-5.68076455461]{3}
		\Vertex[L=$$,x=-7.41807648051,y=-12.0]{4}
		\Vertex[L=$$,x=-8.2282219653,y=-10.5500137213]{5}
		\Vertex[L=$$,x=-8.31965873157,y=-9.43065775746]{6}
		\Vertex[L=$$,x=-2.79099698018,y=-7.23200328455]{7}
		\Vertex[L=$$,x=-3.64711528644,y=0.0]{8}
		\Vertex[L=$$,x=-7.62513565811,y=-4.25517214802]{9}
		\Vertex[L=$$,x=-9.00849770767,y=-10.6119837656]{10}
		\Vertex[L=$$,x=0.0,y=-6.26269082941]{11}
		\Vertex[L=$$,x=-0.736424467365,y=-9.60870243101]{12}
		\Vertex[L=$$,x=-3.60306317538,y=-5.45069027239]{13}
		\Vertex[L=$$,x=-8.16053354447,y=-1.14745909026]{14}
		\Vertex[L=$$,x=-4.96260468894,y=-3.86939556043]{15}
		\Vertex[L=$$,x=-10.4161473923,y=-11.858565547]{16}
		\Vertex[L=$$,x=-4.64880775258,y=-8.44746411056]{17}
		\Vertex[L=$$,x=-4.77091170078,y=-2.03012983611]{18}
		\Vertex[L=$$,x=-6.92548047587,y=-5.18668714868]{19}
		\Vertex[L=$$,x=-5.10236679612,y=-1.86279621302]{20}
		\Vertex[L=$$,x=-1.97387591258,y=-6.27097416807]{21}
		\Vertex[L=$$,x=-0.60511802572,y=-2.73227798579]{22}
		\Vertex[L=$$,x=-11.6385752427,y=-6.22654295901]{23}
		\Vertex[L=$$,x=-9.15659493275,y=-7.52179277203]{24}
		\Vertex[L=$$,x=-9.75120468561,y=-7.19812654379]{25}
		\Vertex[L=$$,x=-3.97413212744,y=-2.72641064619]{26}
		\Vertex[L=$$,x=-8.8727670449,y=-5.95926390596]{27}
		\Vertex[L=$$,x=-8.82273639963,y=-3.26041717989]{28}
		\Vertex[L=$$,x=-10.9973433828,y=-1.68877652464]{29}
		\Vertex[L=$$,x=-4.72186545123,y=-2.57769968416]{30}
		\Vertex[L=$$,x=-12.0,y=-5.93854424321]{31}
		\Vertex[L=$$,x=-7.63107150986,y=-1.62257897189]{32}
		\Vertex[L=$$,x=-8.64240356967,y=-3.7683416193]{33}
		\AddVertexColor{white!90, draw=black}{0,1,2,3,4,5,6,7,10,11,12,13,16,17,19,21}
		\AddVertexColor{blue!90, draw=black}{8,9,14,15,18,20,22,23,24,25,26,27,28,29,30,31,32,33}
		\tikzstyle{EdgeStyle}=[line width=0.7pt, opacity = 0.3]
		\tikzstyle{LabelStyle}=[fill=white]
		\Edge[label=](0)(1)
		\Edge[label=](0)(2)
		\Edge[label=](0)(3)
		\Edge[label=](0)(4)
		\Edge[label=](0)(5)
		\Edge[label=](0)(6)
		\Edge[label=](0)(7)
		\Edge[label=](0)(8)
		\Edge[label=](0)(10)
		\Edge[label=](0)(11)
		\Edge[label=](0)(12)
		\Edge[label=](0)(13)
		\Edge[label=](0)(17)
		\Edge[label=](0)(19)
		\Edge[label=](0)(21)
		\Edge[label=](0)(31)
		\Edge[label=](1)(2)
		\Edge[label=](1)(3)
		\Edge[label=](1)(7)
		\Edge[label=](1)(13)
		\Edge[label=](1)(17)
		\Edge[label=](1)(19)
		\Edge[label=](1)(21)
		\Edge[label=](1)(30)
		\Edge[label=](2)(3)
		\Edge[label=](2)(32)
		\Edge[label=](2)(7)
		\Edge[label=](2)(8)
		\Edge[label=](2)(9)
		\Edge[label=](2)(13)
		\Edge[label=](2)(27)
		\Edge[label=](2)(28)
		\Edge[label=](3)(7)
		\Edge[label=](3)(12)
		\Edge[label=](3)(13)
		\Edge[label=](4)(10)
		\Edge[label=](4)(6)
		\Edge[label=](5)(16)
		\Edge[label=](5)(10)
		\Edge[label=](5)(6)
		\Edge[label=](6)(16)
		\Edge[label=](8)(32)
		\Edge[label=](8)(30)
		\Edge[label=](8)(33)
		\Edge[label=](9)(33)
		\Edge[label=](13)(33)
		\Edge[label=](14)(32)
		\Edge[label=](14)(33)
		\Edge[label=](15)(32)
		\Edge[label=](15)(33)
		\Edge[label=](18)(32)
		\Edge[label=](18)(33)
		\Edge[label=](19)(33)
		\Edge[label=](20)(32)
		\Edge[label=](20)(33)
		\Edge[label=](22)(32)
		\Edge[label=](22)(33)
		\Edge[label=](23)(32)
		\Edge[label=](23)(25)
		\Edge[label=](23)(27)
		\Edge[label=](23)(29)
		\Edge[label=](23)(33)
		\Edge[label=](24)(25)
		\Edge[label=](24)(27)
		\Edge[label=](24)(31)
		\Edge[label=](25)(31)
		\Edge[label=](26)(33)
		\Edge[label=](26)(29)
		\Edge[label=](27)(33)
		\Edge[label=](28)(33)
		\Edge[label=](28)(31)
		\Edge[label=](29)(32)
		\Edge[label=](29)(33)
		\Edge[label=](30)(32)
		\Edge[label=](30)(33)
		\Edge[label=](31)(32)
		\Edge[label=](31)(33)
		\Edge[label=](32)(33)
		\end{tikzpicture}}
	\hspace{20pt}
	\subfloat[GEMSEC]{
		\begin{tikzpicture}[scale=0.3,transform shape]
		\tikzstyle{VertexStyle}=[minimum size = 15pt,inner sep=0pt, shape = circle]
		\Vertex[L=$$,x=-1.56426748103,y=-8.02644495193]{0}
		\Vertex[L=$$,x=-1.64972259019,y=-6.27978860133]{1}
		\Vertex[L=$$,x=-2.84156213579,y=-6.0633432642]{2}
		\Vertex[L=$$,x=-1.03876941173,y=-6.2607369678]{3}
		\Vertex[L=$$,x=-1.13050368569,y=-10.1922479378]{4}
		\Vertex[L=$$,x=-1.292833861,y=-10.4674319486]{5}
		\Vertex[L=$$,x=-1.17178420358,y=-10.779831986]{6}
		\Vertex[L=$$,x=-1.11292495714,y=-6.26776410251]{7}
		\Vertex[L=$$,x=-8.31704326426,y=-1.71856709387]{8}
		\Vertex[L=$$,x=-8.91665241298,y=-0.69013309267]{9}
		\Vertex[L=$$,x=-0.944238038084,y=-9.88652135393]{10}
		\Vertex[L=$$,x=-0.463922613411,y=-8.20593635368]{11}
		\Vertex[L=$$,x=0.0,y=-7.53628115869]{12}
		\Vertex[L=$$,x=-1.89143495637,y=-5.61552349455]{13}
		\Vertex[L=$$,x=-11.2121218691,y=-0.528274439662]{14}
		\Vertex[L=$$,x=-10.6436762805,y=0.0]{15}
		\Vertex[L=$$,x=-1.54945463139,y=-12.0]{16}
		\Vertex[L=$$,x=-0.645163604667,y=-6.61514648118]{17}
		\Vertex[L=$$,x=-11.6188782549,y=-0.56262298113]{18}
		\Vertex[L=$$,x=-2.23588692788,y=-6.18535784683]{19}
		\Vertex[L=$$,x=-10.8772830652,y=-0.062786724843]{20}
		\Vertex[L=$$,x=-0.831268890897,y=-7.55453254404]{21}
		\Vertex[L=$$,x=-11.1015893633,y=-0.216519956195]{22}
		\Vertex[L=$$,x=-12.0,y=-1.07519334371]{23}
		\Vertex[L=$$,x=-11.7832813019,y=-3.12993424057]{24}
		\Vertex[L=$$,x=-11.8370004697,y=-2.79183803921]{25}
		\Vertex[L=$$,x=-11.1044498507,y=-0.570444115495]{26}
		\Vertex[L=$$,x=-10.5152561495,y=-1.95565655421]{27}
		\Vertex[L=$$,x=-9.41573059132,y=-1.36779780215]{28}
		\Vertex[L=$$,x=-11.5974272012,y=-0.941871133535]{29}
		\Vertex[L=$$,x=-8.77671251696,y=-1.02866749943]{30}
		\Vertex[L=$$,x=-10.1288211921,y=-2.3658009608]{31}
		\Vertex[L=$$,x=-10.7463572286,y=-1.79613605322]{32}
		\Vertex[L=$$,x=-10.4208955442,y=-1.86834459039]{33}
		\AddVertexColor{white, draw=black}{0,1,2,3,4,5,6,7,10,11,12,13,16,17,19,21}
		\AddVertexColor{blue!90, draw=black}{8,9,14,15,18,20,22,23,24,25,26,27,28,29,30,31,32,33}
		\tikzstyle{EdgeStyle}=[line width=0.7pt, opacity = 0.3]
		\tikzstyle{LabelStyle}=[fill=white]
		\Edge[label=](0)(1)
		\Edge[label=](0)(2)
		\Edge[label=](0)(3)
		\Edge[label=](0)(4)
		\Edge[label=](0)(5)
		\Edge[label=](0)(6)
		\Edge[label=](0)(7)
		\Edge[label=](0)(8)
		\Edge[label=](0)(10)
		\Edge[label=](0)(11)
		\Edge[label=](0)(12)
		\Edge[label=](0)(13)
		\Edge[label=](0)(17)
		\Edge[label=](0)(19)
		\Edge[label=](0)(21)
		\Edge[label=](0)(31)
		\Edge[label=](1)(2)
		\Edge[label=](1)(3)
		\Edge[label=](1)(7)
		\Edge[label=](1)(13)
		\Edge[label=](1)(17)
		\Edge[label=](1)(19)
		\Edge[label=](1)(21)
		\Edge[label=](1)(30)
		\Edge[label=](2)(3)
		\Edge[label=](2)(32)
		\Edge[label=](2)(7)
		\Edge[label=](2)(8)
		\Edge[label=](2)(9)
		\Edge[label=](2)(13)
		\Edge[label=](2)(27)
		\Edge[label=](2)(28)
		\Edge[label=](3)(7)
		\Edge[label=](3)(12)
		\Edge[label=](3)(13)
		\Edge[label=](4)(10)
		\Edge[label=](4)(6)
		\Edge[label=](5)(16)
		\Edge[label=](5)(10)
		\Edge[label=](5)(6)
		\Edge[label=](6)(16)
		\Edge[label=](8)(32)
		\Edge[label=](8)(30)
		\Edge[label=](8)(33)
		\Edge[label=](9)(33)
		\Edge[label=](13)(33)
		\Edge[label=](14)(32)
		\Edge[label=](14)(33)
		\Edge[label=](15)(32)
		\Edge[label=](15)(33)
		\Edge[label=](18)(32)
		\Edge[label=](18)(33)
		\Edge[label=](19)(33)
		\Edge[label=](20)(32)
		\Edge[label=](20)(33)
		\Edge[label=](22)(32)
		\Edge[label=](22)(33)
		\Edge[label=](23)(32)
		\Edge[label=](23)(25)
		\Edge[label=](23)(27)
		\Edge[label=](23)(29)
		\Edge[label=](23)(33)
		\Edge[label=](24)(25)
		\Edge[label=](24)(27)
		\Edge[label=](24)(31)
		\Edge[label=](25)(31)
		\Edge[label=](26)(33)
		\Edge[label=](26)(29)
		\Edge[label=](27)(33)
		\Edge[label=](28)(33)
		\Edge[label=](28)(31)
		\Edge[label=](29)(32)
		\Edge[label=](29)(33)
		\Edge[label=](30)(32)
		\Edge[label=](30)(33)
		\Edge[label=](31)(32)
		\Edge[label=](31)(33)
		\Edge[label=](32)(33)
		\end{tikzpicture}
		
	}	

	\caption{Zachary's Karate club graph~\cite{zachary1977information}. White nodes: instructor's group; blue nodes: president's group. GEMSEC produces embedding with more tightly clustered communities.}\label{fig:zachary}
\end{figure}
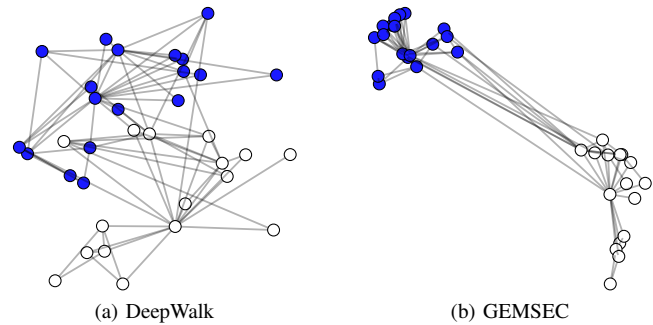

While these methods preserve the proximity of nodes in the graph sense, they do not have an explicit preference for preserving social communities. Thus, in this paper, we develop a machine learning approach that considers clustering when embedding the network and includes a parameter to control the closeness of nodes in the same community. Figure~\ref{fig:zachary}(a) shows the embedding obtained by the standard \textit{Deepwalk} method, where communities are coherent, but not clearly separated in the embedding. The method described in this paper, called \textit{GEMSEC}, is able to produce clusters that are tightly embedded and separated from each other (Fig.~\ref{fig:zachary}(b)).

\subsection{Our Contributions}
\textit{GEMSEC} is an algorithm that considers the two problems of embedding and community detection simultaneously, and as a result, the two solutions of embedding and clustering can inform and improve each other. Through iterations, the embedding converges toward one where nodes are placed close to their neighbors in the network, while at the same time clusters in the embedding space are well separated. 
%The algorithm is based on the sequence-based embedding paradigm. In this approach, sequences of neighboring nodes are extracted using a sampling procedure such as a random walk. Nodes that frequently occur together are then considered ``close''  and embedded near each other. In this optimization, 

The algorithm is based on the paradigm of sequence-based node embedding procedures that create $d$ dimensional feature representations of nodes in an abstract feature space. Sequence-based node embeddings embed pairs of nodes close to each other if they occur frequently within a small window of each other in a random walk. This problem can be formulated as minimizing the negative log-likelihood of observed neighborhood samples (Sec.~\ref{sec:model}) and is called the skip-gram optimization~\cite{mikolov_1}. We extend this objective function to include a clustering cost. The formal description is presented in Subsection~\ref{sec:cluster}. The resulting optimization problem is solved with a variant of mini-batch gradient descent~\cite{adampaper}. 

The detailed algorithm is presented in Subsection~\ref{sec:algorithm}. %Simply, our approach creates a $d$ dimensional representation for each node as well as a set $C$ of cluster centers in the embedding space. 
%Our model is agnostic of the way neighbors of vertices are sampled. 
%In our experiments we used first and second-order truncated random walks \cite{deepwalk,grover2016node2vec}.
%The fundamental idea in our work is that nodes in the same community are likely to have similar neighborhoods, while nodes with similar neighborhoods should be embedded close to each other. We directly exploit this by adding a cost element which enforces the created representation to be clustered. 
%Looking at Figure \ref{fig:zachary} the intuition behind our algorithm can be easily understood. 
By enforcing clustering on the embedding, \textit{GEMSEC} reveals the natural community structure (e.g. Figure \ref{fig:zachary}).Our approach improves over existing methods of simultaneous embedding and clustering~\cite{cavallari2017learning,wang2017community, ye2018deep} and shows that community sensitivity can be directly incorporated into the skip-gram style optimization to obtain greater accuracy and efficiency.

In social networks, nodes in the same community tend to have similar groups of friends, which is expressed as high neighborhood overlap. This fact can be leveraged to produce clusters that are better aligned with the underlying communities. We achieve this effect using a regularization procedure -- a smoothness regularization added to the basic optimization achieves more coherent community detection. The effect can be seen in Figure~\ref{fig:regularization}, where a somewhat uncertain community affiliation suggested by the randomized sampling is sharpened by the smoothness regularization. This technique is described in Subsection~\ref{sec:reglularization}.

In experimental evaluation %of \textit{GEMSEC} primarily focuses on non-overlapping community detection and node-labeling. 
we demonstrate that  \textit{GEMSEC} outperforms -- in clustering quality -- the state of the art neighborhood based~\cite{deepwalk,grover2016node2vec}, multi-scale~\cite{tang2015line,perozzidontwalk} and community aware embedding methods~\cite{cavallari2017learning,wang2017community,ye2018deep}. 
%In terms of the clusterings' modularity, the performance advantage of our proposed method over the baseline community detection procedures varies between 0.93\% and 3.44\%. 
We present new social datasets from the streaming service Deezer and show that the clustering can improve music recommendations.
%In terms of micro-averaged $F_1$ score, the performance increase compared to the baselines is between 3.02\% and 4.95\% on the task of predicting genres liked by users using the embedding features. 
The clustering performance of \textit{GEMSEC} is found to be robust to hyperparameter changes, and the runtime complexity of our method is linear in the size of the graphs.

To summarize, the main contributions of our work are:
\begin{enumerate}
	\item \textit{GEMSEC}: a sequence sampling-based learning model which learns an embedding of the nodes at the same time as it learns a clustering of the nodes.
	\item Clustering in \textit{GEMSEC} can be aligned to network neighborhoods by a smoothness regularization added to the optimization. This enhances the algorithm's sensitivity to natural communities.
	\item Two new large social network datasets are introduced -- from Facebook and Deezer data.
	\item Experimental results show that the embedding process runs linearly in the input size. It generally performs well in quality of embedding and in particular outperforms existing methods on cluster quality measured by modularity and subsequent recommendation tasks.
\end{enumerate}

We start with reviewing related work in the area and relation to our approach in the next section.  A high-performance Tensorflow reference implementation of \textit{GEMSEC}  and the datasets that we collected can be accessed online\footnote{https://github.com/benedekrozemberczki/GEMSEC}.

\section{Related Work}\label{sec:related_work}
There is a long line of research in metric embedding -- for example, embedding discrete metrics into trees~\cite{fakcharoenphol2004tight} and into vector spaces~\cite{matouvsek2002lectures}. Optimization-based representation of networks has been used for routing and navigation in domains such as sensor networks and robotics~\cite{yu2011spherical,huang2014bounded}. Representations in hyperbolic spaces have emerged as a technique to preserve richer network structures~\cite{sarkar2011low,de2018representation,zeng2010resilient}. 

Recent advances in node embedding procedures have made it possible to learn vector features for large real-world graphs~\cite{deepwalk,tang2015line,grover2016node2vec}. Features extracted with these \textit{sequence-based node embedding} procedures can be used for predicting social network users' missing age~\cite{goyal2017graph}, the category of scientific papers in citation networks~\cite{perozzidontwalk} and the function of proteins in protein-protein interaction networks~\cite{grover2016node2vec}. Besides supervised learning tasks on nodes the extracted features can be used for graph visualization~\cite{goyal2017graph}, link prediction~\cite{grover2016node2vec} and community detection~\cite{cavallari2017learning}. %Sequence based node embedding procedures were inspired by distributed word representations, specifically by the skip-gram model~\cite{mikolov_1}. %In this model representation of words are learned to minimize the negative log-likelihood of observing words in the neighborhood (context) of a specific word. These models use sentences in order to generate the contexts which one can think of as linear sequences of strings. The minimization problem itself is solved with stochastic gradient descent and made tractable by the use of negative sampling \cite{gutmann2010noise}. In a similar manner, node embedding models first sample sequences of vertices and extract noisy proximity statistics from these sequences within a fixed context. Co-occurrence statistics are encoded with a single hidden layer neural network and node-specific weights in the input layer are used as vertex representations.

Sequence based embedding commonly considers variations in the sampling strategy that is used to obtain vertex sequences --  truncated random walks being the simplest strategy~\cite{deepwalk}. More involved methods include second-order random walks~\cite{grover2016node2vec}, skips in random walks~\cite{perozzidontwalk} and diffusion graphs~\cite{rozemberczki2018fast}. It is worth noting that these models implicitly approximate matrix factorizations for different matrices that are expensive to factorize explicitly \cite{qiu2018network}. %Changing the sampling strategy corresponds to changing the matrix that is being factorized.

Our work extends the literature of node embedding algorithms which are community aware. Earlier works in this category did not directly extend the skip-gram embedding framework. \textit{M-NMF}~\cite{wang2017community} applies computationally expensive non-negative matrix factorization with a modularity constraint term. The procedure \textit{DANMF} \cite{ye2018deep} uses hierarchical non-negative matrix factorization to create community-aware node embeddings. % to a matrix which is a sum of the neighborhood overlap and adjacency matrices. This procedure is similar to \textit{GEMSEC} as it gives a fixed number of clusters, but differs as it encodes first and second-order proximity. 
\textit{ComE} \cite{cavallari2017learning} is a more scalable approach, but it assumes that in the embedding space the communities fit a gaussian structure, and aims to model them by a mixture of Gaussians. In comparison to these methods, \textit{GEMSEC} provides greater control over community sensitivity of the embedding process, it is independent of the specific neighborhood sampling methods and is computationally efficient. 

\section{Graph Embedding with Self Clustering}\label{sec:model}

For a graph $G=(V,E)$, a node embedding is a mapping $f:V\to \mathbb{R}^d$ where $d$ is the dimensionality of the embedding space. For each node $v \in V$ we create a $d$ dimensional representation. Alternatively, the embedding $f$ is a $|V| \times d$ real-valued matrix. In sequence-based embedding, sequences of neighboring nodes are sampled from the graph. Within a sequence, a node $v$ occurs in the {\em context} of a window $\omega$ within the sequence. Given a sample $S$ of sequences, we refer to the collection of windows containing $v$ as $N_{S}(v)$. Earlier works have proposed random walks, second-order random walks or branching processes to obtain $N_S(v)$. In our experiments, we used unweighted first and second-order random walks for node sampling~\cite{deepwalk,grover2016node2vec}.

Our goal is to minimize the negative log-likelihood of observing neighborhoods of source nodes conditional on feature vectors that describe the position of nodes in the embedding space. Formally, the optimization objective is: 
\begin{equation}{\small
\min_{f}\quad \sum_{v \in V}- \log P(N_S(v)|f(v))\label{eq:negative_loglikelihood}}
\end{equation}
for a suitable probability function $P(\cdot|\cdot)$. To define this $P$, we consider two standard properties (see~\cite{grover2016node2vec}) expected of the embedding $f$ in relation to $N_{S}$. First, it should be possible to factorize $P(N_S(v)|f(v))$ in line with \textit{conditional independence} with respect to $f(v)$. Formally:
\begin{align}{\small
P(N_S(v)|f(v)) = \prod\limits_{n_i \in N_S(v)}P(n_i \in N_S(v)\mid f(v),f(n_i)).\label{eq:condi}}
\end{align} 
Second, it should satisfy \textit{symmetry in the feature space}, meaning that source and neighboring nodes have a symmetric effect on each other in the embedding space. A softmax function on the pairwise dot products of node representations with $f(v)$ to get $P(n_i \in N_S(v)\mid f(v),f(n_i))$ express such a property:

\begin{align}{\small
P(n_i \in N_S(v)\mid f(v),f(n_i)) = \frac{\exp (f(n_i)\cdot f(v))}{\sum\limits_{u \in V}\exp(f(u)\cdot (f(v))}.\label{eq:dotprod}}
\end{align} 
Substituting \eqref{eq:condi} and \eqref{eq:dotprod} into the optimization function, we get: 
\begin{small}
\begin{align}
 \min_{f}&\sum\limits_{v \in V} \left [\ln \left(\smashoperator{\sum\limits_{u\in V}} \exp(f(v)\cdot f(u))\right)-\smashoperator{\sum\limits_{n_i\in N_S(v)}}f(n_i)\cdot f(v) \right].\label{eq:classic_opti}
\end{align}
\end{small}
The partition function in Equation \eqref{eq:classic_opti} enforces nodes to be embedded in a low volume space around the origin, while the second term forces nodes with similar sampled neighborhoods to be embedded close to each other.
\subsection{Learning to Cluster}\label{sec:cluster} 
Next, we extend the optimization to pay attention to the clusters it forms.  
We include a clustering cost similar to $k$-means, measuring the distance from nodes to their cluster centers. This augmented optimization problem is described by minimizing a loss function over the embedding $f$ and position of cluster centers $\mu$, that is, $\dst\min_{f,\mu}\mathcal{L}$, where: 

\begin{small}
\begin{align}
\mathcal{L} = &\quad \underbrace{\sum\limits_{v \in V} \left [\ln \left(\sum\limits_{u\in V} \exp(f(v)\cdot f(u))\right)-\sum\limits_{n_i\in N_S(v)}f(n_i)\cdot f(v) \right]}_{\text{Embedding cost}}\nonumber\\
&\quad +\underbrace{\gamma \cdot \sum_{v \in V} \min_{c\in C} \left \|f(v)-\mu_c  \right \|_2}_{\text{Clustering cost}}.\label{eq:proper_opti}
\end{align}
\end{small}

In Equation \eqref{eq:proper_opti} we have $C$ the set of cluster centers -- the $c^{th}$ cluster mean is denoted by $\mu_c$. Each of these cluster centers is a $d$-dimensional vector in the embedding space. The idea is to minimize the distance from each node to its nearest cluster center. The weight coefficient of the clustering cost is given by the hyperparameter $\gamma$.  Evaluating the partition function in the proposed objective function for all of the source nodes has a $\mathcal{O}(|V|^2)$ runtime complexity. Because of this, we approximate the partition function term with negative sampling which is a form of noise contrastive estimation \cite{mikolov_1,gutmann2010noise}. 
\begin{figure}[h!]
	\centering
	\subfloat[Node capture.\label{nodecapture}]{
		\begin{tikzpicture}[scale=0.35,transform shape]
		\tikzstyle{VertexStyle}=[minimum size = 15pt,inner sep=0pt, shape = circle]
		\draw [dashed,line width=1.7pt, opacity = 0.45] (4.5,-2.5) -- (3.5,2.5);
		\Vertex[L=$$,x=0,y=-1.5]{0}
		\Vertex[L=$$,x=0,y=1.5]{1}
		\Vertex[L=$$,x=3,y=0]{2}
		\Vertex[L=$$,x=6,y=0]{3}
		\Vertex[L=$$,x=7,y=1.5]{4}
		\Vertex[L=$$,x=10,y=-1.5]{5}
		\Vertex[L=$$,x=10,y=1.5]{6}
		\Vertex[L=$$,x=7,y=-1.5]{7}
		\AddVertexColor{white!90, draw=black}{0,1,2,3}
		\AddVertexColor{blue!90, draw=black}{4,5,6,7}
		\tikzstyle{EdgeStyle}=[line width=0.7pt, opacity = 0.45]
		\tikzstyle{LabelStyle}=[fill=white]
		\Edge[label=](0)(1)
		\Edge[label=](1)(2)
		\Edge[label=](2)(0)
		\Edge[label=](0)(3)
		\Edge[label=](1)(3)
		\Edge[label=](2)(3)
		\Edge[label=](4)(3)
		\Edge[label=](7)(3)

		\Edge[label=](5)(6)
		\Edge[label=](7)(6)
		\Edge[label=](7)(5)
		\Edge[label=](4)(6)
		\Edge[label=](4)(7)
		\Edge[label=](4)(5)
		
		\end{tikzpicture}

	}
	\hspace{10pt}
	\subfloat[Empty initialization.\label{wronginit}]{		\begin{tikzpicture}[scale=0.35,transform shape]
		\tikzstyle{VertexStyle}=[minimum size = 15pt,inner sep=0pt, shape = circle]
		\draw [dashed,line width=1.7pt, opacity = 0.45] (-1,-1) -- (-2,4);
		\Vertex[L=$$,x=0,y=0]{0}
		\Vertex[L=$$,x=3,y=0]{1}
		\Vertex[L=$$,x=3,y=3]{2}
		\Vertex[L=$$,x=0,y=3]{3}
		
		\Vertex[L=$$,x=6,y=0]{4}
		\Vertex[L=$$,x=9,y=3]{5}
		\Vertex[L=$$,x=9,y=0]{6}
		\Vertex[L=$$,x=6,y=3]{7}
		\AddVertexColor{white!90, draw=black}{0,1,2,3}
		\AddVertexColor{blue!90, draw=black}{4,5,6,7}
		\tikzstyle{EdgeStyle}=[line width=0.7pt, opacity = 0.45]
		\tikzstyle{LabelStyle}=[fill=white]
		\Edge[label=](0)(1)
		\Edge[label=](1)(2)
		\Edge[label=](2)(3)
		\Edge[label=](1)(3)
		\Edge[label=](2)(0)
		\Edge[label=](3)(0)
		
		\Edge[label=](1)(7)
		
		\Edge[label=](5)(6)
		\Edge[label=](7)(6)
		\Edge[label=](7)(5)
		\Edge[label=](4)(6)
		\Edge[label=](4)(7)
		\Edge[label=](4)(5)
		
		\end{tikzpicture}}
	
	\caption[Potential issues with cluster cost weighting and cluster initialization.]{Potential issues with cluster cost weighting and cluster initialization. Different node colors denote different ground truth community memberships and the computed cluster boundary is denoted by the dashed line. In Subfigure \ref{nodecapture} a single white node is captured in a cluster with the blue nodes due to clustering weight $\gamma$ being high. In Subfigure \ref{wronginit} an empty cluster is initialized with no nodes in it. It is plausible that the cluster center remains empty throughout the optimization process.}
	
\end{figure}
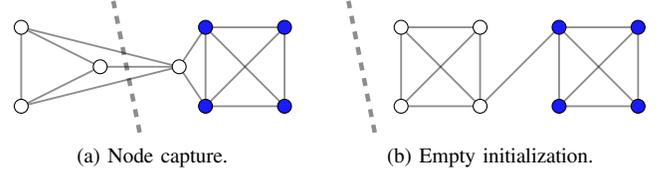
\begin{align}
\frac{\partial \mathcal{L}}{\partial f(v^{\ast})}=&\underbrace{\frac{\sum\limits_{u\in V} \exp(f(v^{\ast})\cdot f(u))\cdot f(u)}{\sum\limits_{u\in V} \exp(f(v^{\ast})\cdot f(u))}}_{\text{Partition function gradient}}-
\underbrace{\sum\limits_{n_i\in N_S(v^\ast)}f(n_i)}_{\text{Neighbor direction}}\nonumber \\
&+\underbrace{\gamma\cdot \frac{f(v^\ast)-\mu_c}{\left \|f(v^\ast)-\mu_c  \right \|_2}}_{\text{Closest cluster direction}}\label{eq:feature_gradient}
\end{align}
The gradients of the loss function in Equation \ref{eq:proper_opti} are important in solving the minimization problem. As a result we can obtain the gradients for node representations and cluster centers. Examining in more detail, the gradient of the objective function $\mathcal{L}$ with respect to the representation of node $v^{\ast} \in V$ is described by Equation \eqref{eq:feature_gradient} if $\mu_c$ is the closest cluster center to $f(v^{\ast})$.

The gradient of the partition function pulls the representation of $v^{\ast}$ towards the origin. The second term moves the representation of $v^{\ast}$ closer to the representations of its neighbors in the embedding space while the third term moves the node closer to the closest cluster center. If we set a high $\gamma$ value the third term dominates the gradient. This will cause the node to gravitate towards the closest cluster center which might not contain the neighbors of $v^{\ast}$. An example is shown in Figure~\ref{nodecapture}. % If node $B$ is initially captured in the cluster on the left-hand side it might never end up in a cluster with its neighbors on the right-hand side. 
%Note that only the position of the closest cluster center affects the representation of  $f(v^{\ast})$. 
If the set of nodes that belong to cluster center $c$ is $V_c$, then  the gradient of the objective function with respect to $\mu_c$ is described by
\begin{equation}
\frac{\partial \mathcal{L}}{\partial \mu_c}=-\gamma\cdot \sum\limits_{v\in V_c}\frac{f(v)-\mu_c}{\left \|f(v)-\mu_c  \right \|_2}.\label{eq:center_gradient}
\end{equation}
{
	\IncMargin{0.5em}	
	\begin{algorithm}[t]
		\DontPrintSemicolon
		\SetAlgoLined
		\KwData{$\mathcal{G} =(V,E)$ -- Graph to be embedded.\\
			\quad \quad\,\,\,\,\,$N$ -- Number of sequence samples per node.\\
			\quad \quad\,\,\,\,\,$l$ -- Length of sequences.\\
			\quad \quad\,\,\,\,\,$\omega$ -- Context size.\\
			\quad \quad\,\,\,\,\,$d$ -- Number of embedding dimensions.\\
			\quad \quad\,\,\,\,\,$|C|$ -- Number of clusters.\\
			\quad \quad\,\,\,\,\,$k$ -- Number of noise samples.\\
			\quad \quad\,\,\,\,\,$\gamma_0$ -- Initial clustering weight coefficient.\\
			\quad \quad\,\,\,\,\,$\alpha_0, \alpha_F$ -- Initial and final learning rate.\\
		}
		\KwResult{$f(v),\,\, \text{ where }v\in V$\\
			\quad\quad  \quad\,\,\,\,	$ \mu_c, \quad \text{ where } c \in C$}
		\vspace{2mm}
		$\text{ Model}\leftarrow \textbf{ Initialize Model}(|V|,d, |C|)$\;
		$t \leftarrow 0$\;
		\For{n\,\,\text{\upshape in}\,\,1:N}{
			\vspace{2mm}   	
			$\widehat{V}\leftarrow \textbf{ Shuffle}(V)$\;
			\For{$v \,\, \text{\upshape in} \,\, \widehat{V}$}{
				\vspace{2mm}
				$t \leftarrow t+1$\;
				$\gamma \leftarrow \textbf{Update } \gamma\,\,(\gamma_0, t, w,l,N, |V|)$\;
				$\alpha\leftarrow \textbf{ Update }\alpha\,\,(\alpha_0,\alpha_F, t, w,l,N, |V|)$\;
				$\text{Sequence}\leftarrow \textbf{Sample Nodes}(\mathcal{G},v, l)$\;
				$\text{Features} \leftarrow \textbf{Extract Features}(\text{ Sequence},\omega)$\;
				$\text{\textbf{Update Weights}}(\text{Model},\text{Features},\gamma, \alpha,k)$
				
			}
			
		}

		\caption{GEMSEC training procedure}\label{GEMSEC_algo}
	\end{algorithm}
	}

In Equation \ref{eq:center_gradient} we see that the gradient moves the cluster center by the sum of coordinates of nodes in the embedding space that belong to cluster $c$. Second, if a cluster ends up empty it will not be updated as elements of the gradient would be zero. Because of this, cluster centers and embedding weights are initialized with the same uniform distribution. A wrong initialization just like the one with an empty cluster in Subfigure~\ref{wronginit} can affect clustering performance considerably.
\subsection{GEMSEC algorithm}\label{sec:algorithm}

We propose an efficient learning method to create \textit{GEMSEC} embeddings which is described with pseudo-code by Algorithm \ref{GEMSEC_algo}. The main idea behind our procedure is the following. To avoid the clustering cost overpowering the graph information (as in Fig.~\ref{nodecapture}), we initialize the system with a low weight $\gamma_{0}\in [0,1]$ for clustering, and through iterations anneal it to $1$. 

The embedding computation proceeds as follows. The weights in the model are initialized based on the number of vertices, embedding dimensions and clusters.  After this, the algorithm makes $N$ sampling repetitions in order to generate vertex sequences from every source node. Before starting a sampling epoch, it shuffles the set of vertices. We set the clustering cost coefficient $\gamma$ (line 7) according to an exponential annealing rule described by Equation \eqref{eq:exp_anneal}.  The learning rate is set to $\alpha$ (line 8) with a linear annealing rule (Equation \eqref{eq:lin_anneal}).

\begin{align}
\gamma&=\gamma_0 \cdot \left(10^{\frac{-t\cdot \log_{10} \gamma_0}{w \cdot l \cdot |V|\cdot N}} \right)\label{eq:exp_anneal}\\
\alpha &=\alpha_0-(\alpha_0-\alpha_F)\cdot \frac{t}{w \cdot l \cdot |V|\cdot N}\label{eq:lin_anneal}
\end{align}

The sampling process reads sequences of length $l$ (line 9) and extracts features using the context window size $\omega$ (line 10). The extracted features, gradient, current learning rate and clustering cost coefficient determine the update to model weights by the optimizer (line 11). In the implementation we utilized a variant of stochastic gradient descent -- the \textit{Adam} optimizer \cite{adampaper}. We approximate the first cost term with noise contrastive estimation to make the gradient descent tractable, drawing $k$ noise samples for each positive sample. If the node sampling is done by first-order random walks the runtime complexity of this procedure will be $\mathcal{O}((\omega\cdot k + |C|)\cdot l\cdot d\cdot |V|\cdot N)$ while \textit{DeepWalk} with noise contrastive estimation has a $\mathcal{O}(\omega\cdot k\cdot l\cdot d\cdot |V|\cdot N)$ runtime complexity.

\subsection{Smoothness Regularization for coherent community detection}\label{sec:reglularization}
%\rik{Notes: Regularization reduces the complexity of the hypothesis. Structural risk minimization. Stabilization -- slight change does not produce a big change. Stable rules do not overfit. }
We have seen in Subsection~\ref{sec:cluster} that there is a tension between what the clustering objective considers to be clusters and what the real communities are in the underlying social network. We can incorporate additional knowledge of social network communities using a machine learning technique called regularization. 

%Regularization can reduce the structural risk and complexity of the output models -- which can be viewed as finding the simpler, more ``natural''  outputs (see~\cite{shalev2014understanding} for an explanation). 
We observe that social networks have natural local properties such as homophily, strong ties between members of a community, etc. Thus, we can incorporate such social network-specific properties in the form of regularization to find more natural embeddings and clusters. 

This regularization effect can be achieved by adding a term $\Lambda$ to the loss function: 
{\small\begin{align}
	\Lambda & =\lambda \cdot\smashoperator{\sum_{(v,u) \in E_S}} w_{(v,u)}\cdot\left \|f(v)-f(u)  \right \|_2\label{eq:smoother},
	\end{align}}
where the weight function $w$ determines the {\em social network cost} of the embedding with respect to properties of the edges traversed in the sampling. We use the neighborhood overlap of an edge -- defined as the fraction of neighbors common to two nodes of the edge relative to the union of the two neighbor sets\footnote{Neighbor sets $N(a)$ and $N(b)$ of nodes $a$ and $b$, the neighborhood overlap of $(a,b)$ is defined as the Jaccard similarity $\frac{N(a)\cap N(b)}{N(a)\cup N(b)}$.}. In experiments on real data, neighborhood overlap is known to be a strong indicator of the strength of relation between members of a social network~\cite{onnela2007structure}. Thus, by treating neighborhood overlap as the weight $w_{v,u}$ of edge $(v,u)$, we can get effective social network clustering, which is confirmed by experiments in the next section. The coeffeicient $\lambda$ lets us tune the contribution of the social network cost in the embedding process. In experiments, the regularized version of the algorithms is found to be more robust to changes in hyperparameters.

The effect of the regularization can be understood intuitively through an example. For this exposition, let us consider matrix representations of the social network describing closeness of nodes. In fact, other skip-gram style learning processes like~\cite{deepwalk,grover2016node2vec} are known to approximate the factorization of a similarity matrix $M$ such as~\cite{qiu2018network}:
$$M_{u,v} = \log\left(\frac{\text{vol}(G)}{\omega}\sum\limits_{r=1}^{\omega} \frac{\sum \limits_{P\in \mathcal{P}^r_{v,u}} \prod\limits_{a\in P\setminus  \left\{v\right\}}  \frac{1}{\deg(a)}}{\deg(v)}\right)-\log(k)
$$
%The target matrix that is being factorized is approximated by doing respectively first or second order random walks as computing it explicitly would have an $\mathcal{O}(|V|^2)$ runtime and memory complexity for $\omega\geq \text{diam}(G)$. For example, in case of \textit{DeepWalk} an element of the target matrix for nodes $v,u$ takes the form:
where $P^r_{v,u}$ is the set of paths going from $v$ to $u$ with length $r$. Elements of the target matrix $M$ grow with number of paths of length at most $\omega$ between the corresponding nodes. Thus $M$ is intended to represent level of connectivity between nodes in terms of a raw graph feature like number of paths. 

%However, this expression has no explicit weight on edges with high neighborhood overlap, which are important for clustering. To resolve this, the node embedding objective function is augmented by adding a smoothness regularization cost element which encourages learned representations to be similar if two sampled nodes share an edge with high neighborhood overlap. We could add a smoothness regularization both to Equations \eqref{eq:classic_opti} and \eqref{eq:proper_opti}. If $\Lambda$ is added to the objective function described by Equation \eqref{eq:classic_opti} we get the smoothed node embedding model. Later we reference this model as \textit{Smooth DeepWalk}. We also add the regularization to the objective function of \textit{GEMSEC} to define \textit{Smooth GEMSEC}.

\begin{figure*}
\centering
\adjustbox{valign=t}{
	\begingroup
	\captionsetup[subfigure]{width=50pt}
\subfloat[The graph \label{fig:barbell_graph}]{
	\begin{tikzpicture}[scale=0.26,transform shape]
		\tikzstyle{VertexStyle}=[minimum size = 25pt,inner sep=0pt, shape = circle]
\Vertex[L=$$,x=0,y=2]{0}
\Vertex[L=$$,x=-2,y=0]{1}
\Vertex[L=$$,x=-1.5,y=-2]{2}
\Vertex[L=$$,x=1.5,y=-2]{3}
\Vertex[L=$$,x=2,y=0.0]{4}

\Vertex[L=$$,x=2,y=3]{5}
\Vertex[L=$$,x=4,y=3]{6}
\Vertex[L=$$,x=6,y=3]{7}

\Vertex[L=$$,x=8,y=4]{9}
\Vertex[L=$$,x=6,y=6]{8}
\Vertex[L=$$,x=6.5,y=8]{10}
\Vertex[L=$$,x=9.5,y=8]{11}
\Vertex[L=$$,x=10,y=6]{12}

		\AddVertexColor{red!90, draw=black}{0,1,2,3,4,5,6,7,8,9,10,11,12}
		\tikzstyle{EdgeStyle}=[line width=2.0pt, opacity = 0.45]
		\tikzstyle{LabelStyle}=[fill=white]
\Edge[label=](0)(1)
\Edge[label=](0)(2)
\Edge[label=](0)(3)
\Edge[label=](0)(4)
\Edge[label=](1)(2)
\Edge[label=](1)(3)
\Edge[label=](1)(4)
\Edge[label=](2)(3)
\Edge[label=](2)(4)
\Edge[label=](3)(4)

\Edge[label=](0)(5)
\Edge[label=](5)(6)
\Edge[label=](6)(7)
\Edge[label=](7)(9)

\Edge[label=](8)(9)
\Edge[label=](8)(10)
\Edge[label=](8)(11)
\Edge[label=](8)(12)
\Edge[label=](9)(10)
\Edge[label=](9)(11)
\Edge[label=](9)(12)
\Edge[label=](10)(11)
\Edge[label=](10)(12)
\Edge[label=](11)(12)

\end{tikzpicture}

}	\endgroup}
\hspace{30pt}
\adjustbox{valign=t}{
		\begingroup
		\captionsetup[subfigure]{width=100pt}
	\subfloat[Target Matrix of Deepwalk. \label{fig:barbell_target}]{
	\resizebox{0.16\textwidth}{!}{
		\begin{tabular}{*{13}{R}}
0.792 & 1.0 & 1.0 & 1.0 & 0.881 & 0.4 & 0.162 & 0.046 & 0.005 & 0.0 & 0.0 & 0.0 & 0.0\\
1.0 & 0.792 & 1.0 & 1.0 & 0.881 & 0.4 & 0.162 & 0.046 & 0.005 & 0.0 & 0.0 & 0.0 & 0.0\\
1.0 & 1.0 & 0.792 & 1.0 & 0.881 & 0.4 & 0.162 & 0.046 & 0.005 & 0.0 & 0.0 & 0.0 & 0.0\\
1.0 & 1.0 & 1.0 & 0.792 & 0.881 & 0.4 & 0.162 & 0.046 & 0.005 & 0.0 & 0.0 & 0.0 & 0.0\\
0.995 & 0.995 & 0.995 & 0.995 & 0.801 & 1.0 & 0.458 & 0.223 & 0.024 & 0.006 & 0.006 & 0.006 & 0.006\\
0.194 & 0.194 & 0.194 & 0.194 & 0.43 & 0.576 & 1.0 & 0.429 & 0.096 & 0.022 & 0.022 & 0.022 & 0.022\\
0.079 & 0.079 & 0.079 & 0.079 & 0.197 & 1.0 & 0.808 & 1.0 & 0.197 & 0.079 & 0.079 & 0.079 & 0.079\\
0.022 & 0.022 & 0.022 & 0.022 & 0.096 & 0.429 & 1.0 & 0.576 & 0.43 & 0.194 & 0.194 & 0.194 & 0.194\\
0.006 & 0.006 & 0.006 & 0.006 & 0.024 & 0.223 & 0.458 & 1.0 & 0.801 & 0.995 & 0.995 & 0.995 & 0.995\\
0.0 & 0.0 & 0.0 & 0.0 & 0.005 & 0.046 & 0.162 & 0.4 & 0.881 & 0.792 & 1.0 & 1.0 & 1.0\\
0.0 & 0.0 & 0.0 & 0.0 & 0.005 & 0.046 & 0.162 & 0.4 & 0.881 & 1.0 & 0.792 & 1.0 & 1.0\\
0.0 & 0.0 & 0.0 & 0.0 & 0.005 & 0.046 & 0.162 & 0.4 & 0.881 & 1.0 & 1.0 & 0.792 & 1.0\\
0.0 & 0.0 & 0.0 & 0.0 & 0.005 & 0.046 & 0.162 & 0.4 & 0.881 & 1.0 & 1.0 & 1.0 & 0.792\\

		\end{tabular}
}
}\endgroup}
\hspace{30pt}
\adjustbox{valign=t}{
	\begingroup
	\captionsetup[subfigure]{width=100pt}	
	\subfloat[Approximation by Deepwalk.\label{fig:deepwalk_reconstruct}]{
	\resizebox{0.16\textwidth}{!}{
		\begin{tabular}{*{13}{R}}
0.989 & 0.997 & 1.0 & 0.995 & 0.949 & 0.423 & 0.213 & 0.161 & 0.126 & 0.134 & 0.14 & 0.13 & 0.137\\
0.984 & 1.0 & 1.0 & 0.999 & 0.928 & 0.395 & 0.198 & 0.151 & 0.127 & 0.138 & 0.143 & 0.132 & 0.141\\
0.985 & 0.998 & 1.0 & 0.996 & 0.94 & 0.405 & 0.204 & 0.154 & 0.125 & 0.134 & 0.139 & 0.129 & 0.137\\
0.982 & 0.999 & 0.998 & 1.0 & 0.917 & 0.389 & 0.194 & 0.149 & 0.129 & 0.14 & 0.146 & 0.134 & 0.143\\
0.96 & 0.951 & 0.965 & 0.94 & 1.0 & 0.506 & 0.275 & 0.193 & 0.119 & 0.121 & 0.124 & 0.119 & 0.124\\
0.532 & 0.503 & 0.516 & 0.495 & 0.629 & 1.0 & 0.746 & 0.664 & 0.261 & 0.231 & 0.221 & 0.236 & 0.229\\
0.264 & 0.248 & 0.256 & 0.243 & 0.336 & 0.733 & 1.0 & 0.945 & 0.453 & 0.383 & 0.363 & 0.399 & 0.381\\
0.185 & 0.175 & 0.179 & 0.173 & 0.219 & 0.606 & 0.878 & 1.0 & 0.594 & 0.511 & 0.479 & 0.534 & 0.499\\
0.134 & 0.137 & 0.134 & 0.139 & 0.125 & 0.221 & 0.389 & 0.55 & 0.984 & 1.0 & 0.968 & 0.987 & 0.978\\
0.135 & 0.14 & 0.137 & 0.143 & 0.12 & 0.184 & 0.311 & 0.447 & 0.945 & 1.0 & 0.971 & 0.96 & 0.979\\
0.144 & 0.15 & 0.146 & 0.153 & 0.127 & 0.181 & 0.303 & 0.432 & 0.943 & 1.0 & 0.983 & 0.977 & 0.981\\
0.134 & 0.138 & 0.135 & 0.14 & 0.121 & 0.193 & 0.333 & 0.48 & 0.959 & 0.987 & 0.974 & 1.0 & 0.965\\
0.14 & 0.146 & 0.143 & 0.149 & 0.125 & 0.186 & 0.316 & 0.446 & 0.944 & 1.0 & 0.973 & 0.959 & 0.981\\
		\end{tabular}
}
}\endgroup}
\hspace{30pt}
\adjustbox{valign=t}{
	\begingroup
	\captionsetup[subfigure]{width=100pt}	
	\subfloat[Regularized Approximation \label{fig:smooth_deepwalk_reconstruct}]{
	\resizebox{0.16\textwidth}{!}{
		\begin{tabular}{*{13}{R}}
			1.0 & 1.0 & 1.0 & 1.0 & 1.0 & 0 & 0 & 0 & 0 & 0 & 0 & 0 & 0\\
			1.0 & 1.0 & 1.0 & 1.0 & 1.0 & 0 & 0 & 0 & 0 & 0 & 0 & 0 & 0\\
			1.0 & 1.0 & 1.0 & 1.0 & 1.0 & 0 & 0 & 0 & 0 & 0 & 0 & 0 & 0\\
			1.0 & 1.0 & 1.0 & 1.0 & 1.0 & 0 & 0 & 0 & 0 & 0 & 0 & 0 & 0\\
			1.0 & 1.0 & 1.0 & 1.0 & 1.0 & 0.127 & 0 & 0 & 0 & 0 & 0 & 0 & 0\\
			0 & 0 & 0 & 0 & 0.127 & 1.0 & 0.696 & 0 & 0 & 0 & 0 & 0 & 0\\
			0 & 0 & 0 & 0 & 0 & 0.349 & 1.0 & 0.366 & 0 & 0 & 0 & 0 & 0\\
			0 & 0 & 0 & 0 & 0 & 0 & 0.679 & 1.0 & 0.127 & 0 & 0 & 0 & 0\\
			0 & 0 & 0 & 0 & 0 & 0 & 0 & 0.127 & 1.0 & 1.0 & 1.0 & 1.0 & 1.0\\
			0 & 0 & 0 & 0 & 0 & 0 & 0 & 0 & 1.0 & 1.0 & 1.0 & 1.0 & 1.0\\
			0 & 0 & 0 & 0 & 0 & 0 & 0 & 0 & 1.0 & 1.0 & 1.0 & 1.0 & 1.0\\
			0 & 0 & 0 & 0 & 0 & 0 & 0 & 0 & 1.0 & 1.0 & 1.0 & 1.0 & 1.0\\
			0 & 0 & 0 & 0 & 0 & 0 & 0 & 0 & 1.0 & 1.0 & 1.0 & 1.0 & 1.0\\
			
		\end{tabular}
}
}\endgroup}
%\hspace{20pt}
%\adjustbox{valign=t}{\subfloat[GEMSEC Reconstruction\label{fig:gemsec_reconstruct}]{
%	\input{./plots_and_tables/gemsec_tab.tex}
%}}
%\hspace{20pt}
%\adjustbox{valign=t}{\subfloat[Smooth GEMSEC Reconstruction \label{fig:smooth_gemsec_reconstruct}]{
%	\input{./plots_and_tables/smooth_gemsec_tab.tex}
%}}

%
%			\begin{subfigure}[b]{.3\linewidth}
%				\input{dw_tab.tex}
%				\caption{DeepWalk}\label{fig:deepwalk_reconstruct}
%			\end{subfigure}
%\vfill
%\vspace{4mm}
%\begin{subfigure}[b]{.3\linewidth}
%						\input{smooth_tab.tex}
%						\caption{Smooth DeepWalk}\label{fig:smooth_deepwalk_reconstruct}
%\end{subfigure}
%\hfill
%\begin{subfigure}[b]{.3\linewidth}
%\input{gemsec_tab.tex}
%\caption{GEMSEC}\label{fig:gemsec_reconstruct}
%\end{subfigure}
%\hfill
%\begin{subfigure}[b]{.3\linewidth}
%\input{smooth_gemsec_tab.tex}
%\caption{Smooth GEMSEC}\label{fig:smooth_gemsec_reconstruct}
%\end{subfigure}
\caption{An example Barbell graph with the corresponding target matrix factorized (window size of 3) by \textit{DeepWalk} \cite{qiu2018network} and the reconstructed target matrices obtained with standard \textit{DeepWalk} and \textit{Smooth DeepWalk}. Regularized optimization produces more well defined communities. While the standard \textit{DeepWalk} model has less well defined clusters.}\label{fig:regularization}	
\end{figure*}
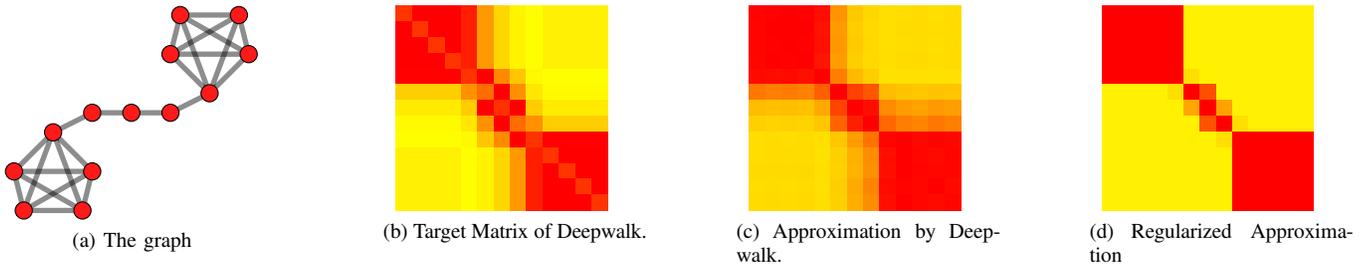
	
%In Equation \eqref{eq:smoother} each sampled edge $(v,u)$ has a regularization weight  $w_{(v,u)}$. The weight can be arbitrarily chosen. One can set weights such as the neighborhood overlap, normalized neighborhood overlap (Jaccard's coefficient) and the minimal set cardinality normalized neighborhood overlap or a simple unit weight on each edge. The hyperparameter $\lambda$ describes the regularization coefficient. If $\lambda$ is high, distant representations of nodes with an edge between them are penalized heavily. The regularization term is being summed over the edges obtained with the sampling strategy. The training procedure described by Algorithm \ref{GEMSEC_algo} needs to be modified slightly to accommodate the introduction of the smoothness regularization term. First, we need to set a regularization coefficient. Using the sampled sequence of vertices we extract the edges on which we want to enforce the regularization term. For each sampled edge $(v,u)$ we have to obtain the weight $w_{v,u}$ and the optimizer besides the regularization coefficient also needs this list of edges and weights for doing an update of model parameters.

The barbell graph in Figure~\ref{fig:barbell_graph} is a typical example with an obvious community structure we can use to analyze the matter. The optimization procedure used by Deepwalk~\cite{deepwalk} aims to converge to a target matrix $M_{u,v}$ shown in Figure~\ref{fig:barbell_target}. Observe that this matrix has fuzzy edges around the communities of the graph, showing a degree of uncertainty. An actual approximation by running the Deepwalk is shown in Figure~\ref{fig:deepwalk_reconstruct}, which naturally incorporates further uncertainty due to sampling. A much more clear output with sharp communities can be obtained by applying a regularized optimization. This can be seen in Figure~\ref{fig:smooth_deepwalk_reconstruct}. 

%\rik{I suggest use ``regularized'' as the main adjective instead of ``smooth''. The regularization effect is the important one here. Smooth regularization is its type. Let me know if there is a reason not to do that.} 

%The smoothness regularization can be explained with an illustrative example. Let us imagine that we have a Barbell graph just like the one depicted in Subfigure \ref{fig:barbell_graph}. The proposed models approximate with truncated random walks a matrix which is a weighted sum of the normalized adjacency matrix powers up to the window size -- just like \textit{DeepWalk} does. This non negative target matrix for the Barbell graph is on Subfigure \ref{fig:barbell_target}. Red elements are large in the target matrix and yellow elements are close to zero. A \textit{DeepWalk} node embedding is essentially obtained by implicit factorization of this target matrix. Using the embedding one can reconstruct the original target matrix as a matrix-matrix product between the embedding and its transpose -- see Subigures \ref{fig:deepwalk_reconstruct} and \ref{fig:smooth_deepwalk_reconstruct}. The effect of smoothness regularization can be understood by looking at these figures. Most noisy values of the target matrix which are between nodes that have low neighborhood overlap are regularized out in the reconstructed target matrix by the enforcement of smoothness.
\setlength{\tabcolsep}{6pt}
\renewcommand{\arraystretch}{1}
\begin{table}[h!]
	\centering
	\caption{Statistics of social networks used in the paper.}
	\label{fig:stats}
		
	{\footnotesize

		\begin{tabular}{lcccc}
			\hline
			\textbf{Source}& \textbf{Dataset}& $|\textbf{V}|$ & $\textbf{Density}$ & \textbf{Transitivity}\\ \hline
			\multirow{8}{*}{Facebook} &  Politicians &   5,908  &0.0024 &  0.3011      \\[0.35em]
			&  Companies &   14,113  & 0.0005    &     0.1532           \\[0.25em]
			&  Athletes &   13,866  & 0.0009  &   0.1292      \\[0.25em]
			&  Media &   27,917  & 0.0005  &   0.1140       \\[0.25em]
			&  Celebrities &   11,565  & 0.0010    & 0.1666           \\[0.25em]
			&  Artists &   50,515  &  0.0006   &    0.1140      \\[0.25em]
			&  Government &   7,057  & 0.0036    &0.2238          \\[0.25em]
			&  TV Shows &   3,892  & 0.0023    & 0.5906        \\[0.25em]\hline
			&Croatia &   54,573  &      0.0004   & 0.1146      \\[0.25em]
			Deezer & Hungary &   47,538  & 0.0002    &      0.0929    \\[0.25em]
			& Romania &   41,773 &  0.0001   &       0.0752  \\[0.35em]
			\hline
	\end{tabular}}

\end{table}

\section{Experimental Evaluation}\label{sec:experiments}
In this section we evaluate the cluster quality obtained by the \textit{GEMSEC} variants, their scalability, robustness and predictive performance on a downstream supervised task.
\begin{table*}[t]

	\caption[Mean modularity of clusterings on the Facebook datasets.]{Mean modularity of clusterings on the Facebook datasets. Each embedding experiment was repeated ten times. Errors in the parentheses correspond to two standard deviations. In terms of modularity $\textit{Smooth GEMSEC}_2$ outperforms the baselines.}
	\label{fig:clust_performance}
	\centering{\footnotesize
		\begin{tabular}{lcccccccc}
			& \textbf{Politicians} & \textbf{Companies} & \textbf{Athletes}   & \textbf{Media}   & \textbf{Celebrities} & \textbf{Artists}& \textbf{Government}& \textbf{TV Shows}   \\
			\hline
			\textbf{Overlap Factorization}            &$\underset{ (\pm 0.008)}{0.810}$&$\underset{( \pm 0.010)}{0.553}$
			&$\underset{( \pm 0.020)}{0.601}$&$\underset{( \pm 0.016)}{0.471}$&$\underset{( \pm 0.01)}{0.551}$ &$\underset{(\pm 0.018)}{0.474}$&$\underset{( \pm 0.024)}{0.608}$ &$\underset{( \pm 0.008)}{0.786}$ \\[0.75em]
			
			\textbf{DeepWalk}&$\underset{(\pm 0.015)}{0.840}$&$\underset{(\pm 0.012)}{0.637}$&$\underset{(\pm 0.012)}{0.649}$&$\underset{(\pm 0.022)}{0.481}$&$\underset{(\pm 0.011)}{0.631}$&$\underset{(\pm 0.029)}{0.508}$&$\underset{(\pm 0.024)}{0.686}$&$\underset{(\pm 0.005)}{0.811}$
			\\[0.75em]
			\textbf{LINE}&$\underset{(\pm 0.014)}{0.841}$&$\underset{(\pm 0.009)}{0.651}$&$\underset{(\pm 0.007)}{0.665}$&$\underset{(\pm 0.012)}{0.558}$&$\underset{(\pm 0.010)}{0.642}$&$\underset{(\pm 0.014)}{0.557}$&$\underset{(\pm 0.017)}{0.690}$&$\underset{(\pm 0.010)}{0.813}$
			\\[0.75em]
			\textbf{Node2Vec}&$\underset{(\pm 0.012)}{0.846}$&$\underset{(\pm 0.008)}{0.664}$&$\underset{(\pm 0.007)}{0.669}$&$\underset{(\pm 0.011)}{0.565}$&$\underset{(\pm 0.013)}{0.643}$&$\underset{(\pm 0.010)}{0.560}$&$\underset{(\pm 0.017)}{0.692}$&$\underset{(\pm 0.016)}{0.827}$
			\\[0.75em]
			\textbf{Walklets}&$\underset{(\pm 0.014)}{0.843}$&$\underset{(\pm 0.012)}{0.655}$&$\underset{(\pm 0.007)}{0.664}$&$\underset{(\pm 0.009)}{0.562}$&$\underset{(\pm 0.043)}{0.621}$&$\underset{(\pm 0.016)}{0.548}$&$\underset{(\pm 0.019)}{0.689}$&$\underset{(\pm 0.015)}{0.819}$\\[0.75em]
			\textbf{ComE}&$\underset{(\pm 0.008)}{0.830}$&$\underset{(\pm 0.005)}{0.654}$&$\underset{(\pm 0.007)}{0.665}$&$\underset{(\pm 0.005)}{\textbf{0.573}}$&$\underset{(\pm 0.010)}{0.635}$&$\underset{(\pm 0.011)}{0.560}$&$\underset{(\pm 0.010)}{0.696}$&$\underset{(\pm 0.011)}{0.806}$
			\\[0.75em]
			\textbf{M-NMF}&$\underset{(\pm 0.014)}{0.816}$&$\underset{(\pm 0.007)}{0.646}$&$\underset{(\pm 0.008)}{0.655}$&$\underset{(\pm 0.004)}{0.561}$&$\underset{(\pm 0.006)}{0.628}$&$\underset{(\pm 0.021)}{0.535}$&$\underset{(\pm 0.011)}{0.668}$&$\underset{(\pm 0.008)}{0.813}$
			\\
			[0.75em]
						\textbf{DANMF}&$\underset{(\pm 0.020)}{0.810}$&$\underset{(\pm 0.005)}{0.648}$&$\underset{(\pm 0.009)}{0.650}$&$\underset{(\pm 0.006)}{0.560}$&$\underset{(\pm 0.011)}{0.628}$&$\underset{(\pm 0.019)}{0.532}$&$\underset{(\pm 0.015)}{0.673}$&$\underset{(\pm 0.014)}{0.812}$
						\\[0.75em]\hline
			\textbf{Smooth DeepWalk}&$\underset{(\pm 0.017)}{0.849}$&$\underset{(\pm 0.007)}{0.667}$&$\underset{(\pm 0.007)}{0.669}$&$\underset{(\pm 0.006)}{0.541}$&$\underset{(\pm 0.008)}{0.643}$&$\underset{(\pm 0.020)}{0.523}$&$\underset{(\pm 0.008)}{0.707}$&$\underset{(\pm 0.008)}{0.835}$\\[0.75em]
			\textbf{GEMSEC}&$\underset{(\pm 0.009)}{0.851}$&$\underset{(\pm 0.013)}{0.662}$&$\underset{(\pm 0.009)}{0.674}$&$\underset{(\pm 0.011)}{0.536}$&$\underset{(\pm 0.014)}{0.636}$&$\underset{(\pm 0.020)}{0.528}$&$\underset{(\pm 0.020)}{0.705}$&$\underset{(\pm 0.010)}{0.833}$\\[0.75em]
			\textbf{Smooth GEMSEC}&$\underset{(\pm 0.006)}{0.855}$&$\underset{(\pm 0.009)}{0.683}$&$\underset{(\pm 0.009)}{\textbf{0.692}}$&$\underset{(\pm 0.009)}{0.567}$&$\underset{(\pm 0.008)}{\textbf{0.649}}$&$\underset{(\pm 0.011)}{0.559}$&$\underset{(\pm 0.008)}{0.710}$&$\underset{(\pm 0.004)}{0.841}$\\[0.75em]\hline
			\textbf{GEMSEC}$_2$&$\underset{(\pm 0.010)}{0.852}$&$\underset{(\pm 0.008)}{0.667}$&$\underset{(\pm 0.008)}{0.683}$&$\underset{(\pm 0.008)}{0.551}$&$\underset{(\pm 0.009)}{0.638}$&$\underset{(\pm 0.020)}{\textbf{0.562}}$&$\underset{(\pm 0.010)}{\textbf{0.712}}$&$\underset{(\pm 0.010)}{0.838}$\\[0.75em]
			\textbf{Smooth GEMSEC}$_2$&$\underset{(\pm 0.006)}{\textbf{0.859}}$&$\underset{(\pm 0.009)}{\textbf{0.684}}$&$\underset{(\pm 0.007)}{\textbf{0.692}}$&$\underset{(\pm 0.010)}{0.571}$&$\underset{(\pm 0.011)}{\textbf{0.649}}$&$\underset{(\pm 0.017)}{\textbf{0.562}}$&$\underset{(\pm 0.010)}{\textbf{0.712}}$&$\underset{(\pm 0.006)}{\textbf{0.847}}$\\[0.75em]
			\hline \\[-1ex]
	\end{tabular}}
\end{table*}
 Results show that \textit{GEMSEC} outperforms or is at par with existing methods in all measures.

\subsection{Datasets}
 For the evaluation of \textit{GEMSEC} real-world social network datasets are used which we collected from public APIs specifically for this work. Table \ref{fig:stats} shows these social networks have a variety of size, density, and level of clustering. We used graphs from two sources:
\begin{itemize}
	\item \textit{Facebook page networks:} These graphs represent mutual like networks among verified Facebook pages -- the types of sites included TV shows, politicians, athletes, and artists among others.
	\item \textit{Deezer user-user friendship networks:} We collected friendship networks from the music streaming site Deezer and included $3$ European countries (Croatia, Hungary, and Romania). For each user, we curated the list of genres loved based on the songs liked by the user.
\end{itemize}
\subsection{Standard parameter settings}
A fixed standard parameter setting is used our experiments, and we indicate any deviations. Models using first order random walk sampling strategy are referenced as $\textit{GEMSEC}$ and $\textit{Smooth GEMSEC}$, second order random walk variants are named as $\textit{GEMSEC}_2$ and $\textit{Smooth GEMSEC}_2$. Random walks with length $80$ are used and $5$ truncated random walks per source node were used. Second-order random walk control hyperparameters \cite{grover2016node2vec} \textit{return} and \textit{in-out} were chosen from  $\left \{2^{-2},2^{-1},1,2,4\right\}$. A window size of $5$ is used for features. Each embedding has $16$ dimensions and we extract $20$ cluster centers. A parameter sweep over hyperparameters was used to obtain the highest average modularity. Initial learning rate values are chosen from $\left \{10^{-2},5\cdot 10^{-3},10^{-3}\right\}$ and the final learning rate is chosen from  $\left \{10^{-3},5\cdot 10^{-4},10^{-4}\right\}$. Noise contrastive estimation uses $10$ negative examples. The initial clustering cost coefficient is  chosen from $\left \{10^{-1},10^{-2},10^{-3}\right\}$. The smoothness regularization term's hyperparameter is $0.0625$ and Jaccard's coefficient is the penalty weight.
\subsection{Cluster Quality}
 Using Facebook page networks we evaluate the clustering performance. Cluster quality is evaluated by modularity -- we assume that a node belongs to a single community. Our results are summarized in Table \ref{fig:clust_performance} based on 10 experimental repetitions and errors in parentheses correspond to two standard deviations. The baselines use the hyperparameters from the respective papers. We used 16-dimensional embeddings throughout. The embeddings obtained with non-community-aware methods were clustered after the embedding by $k$-means clustering to extract 20 cluster centers. 
Specifically, comparisons are made with:
\begin{enumerate}
	\item \textit{Overlap Factorization} \cite{ahmed2013distributed}: Factorizes the neighborhood overlap matrix to create features.
	\item \textit{DeepWalk} \cite{deepwalk}: Approximates the sum of the adjacency matrix powers with first order random walks and implicitly factorizes it.
	\item \textit{LINE} \cite{tang2015line}: Implicitly factorizes the sum of the first two powers for the normalized adjacency matrix and the resulting node representation vectors are concatenated together to form a multi-scale  representation.
	\item \textit{Node2vec} \cite{grover2016node2vec}: Factorizes a neighbourhood matrix obtained with second order random walks. The \textit{in-out} and \textit{return} parameters of the second-order random walks were chosen from the $\left\{2^{-2},2^{-1},1,2,4\right\}$ set to maximize modularity.
	\item \textit{Walklets} \cite{perozzidontwalk}: Approximates with first order random walks each adjacency matrix power individually and implicitly factorizes the target matrix. These embeddings are concatenated to form a multi-scale representation of nodes.
	
	\item \textit{ComE} \cite{cavallari2017learning}: Uses a Gaussian mixture model to learn an embedding and clustering jointly using random walk features.
	\item \textit{M-NMF} \cite{wang2017community}: Factorizes a matrix which is a weighted sum of the first two proximity matrices with a modularity based regularization constraint.
	\item \textit{DANMF} \cite{ye2018deep}: Decomposes a weighted sum of the first two proximity matrices hierarchically to obtain cluster memberships with an autoencoder-like non-negative matrix factorization model.
\end{enumerate}
\textit{Smooth GEMSEC}, $\textit{GEMSEC}_2$ and $\textit{Smooth GEMSEC}_2$ consistently outperform the neighborhood conserving node embedding methods and the competing community aware methods. The relative advantage of $\textit{Smooth GEMSEC}_2$ over the benchmarks is highest on the Athletes dataset as the clustering's modularity is 3.44\% higher than the best performing baseline. It is the worst on the Media dataset with a disadvantage of 0.35\% compared to the strongest baseline. Use of smoothness regularization has sometimes non-significant, but definitely positive effect on the clustering performance of \textit{Deepwalk}, \textit{GEMSEC} and $\textit{GEMSEC}_2$.
\input{sensitivity.tex}
\subsection{Sensitivity Analysis for hyperparameters}
We tested the effect of hyperparameter changes to clustering performance. The Politicians Facebook graph is embedded with the standard parameter settings while the initial and final learning rates are set to be $10^{-2}$ and $5\cdot 10^{-3}$ respectively, the clustering cost coefficient is 0.1 and we perturb certain hyperparameters. The second-order random walks used \textit{in-out} and \textit{return} parameters of $4$. In Figure \ref{fig:sensi} each data point represents the mean modularity calculated from $10$ experiments. %We test the sensitivity of clustering performance to cluster center number, context size, dimension number, random walk length, clustering cost coefficient and the number of initiated random walks per source node.
Based on the experimental results we make two observations. First, \textit{GEMSEC} model variants give high-quality clusters for a wide range of parameter settings. Second, introducing smoothness regularization makes \textit{GEMSEC} models more robust to hyperparameter changes. This is particularly apparent across varying the number of clusters. %Specifically, we observe that increasing the cluster number above 30 decreases slightly the average modularity. In the case of the non-smooth models, the decrease in cluster quality is quite considerable.
The length of truncated random walks and the number of random walks per source node above a certain threshold has only a marginal effect on the community detection performance. %Interestingly, we also have empirical evidence for the node capture phenomenon --- if the clustering cost coefficient is too high vanilla \textit{GEMSEC} models have a poor clustering performance. Finally, there is strong evidence that both \textit{GEMSEC} and \textit{Smooth GEMSEC} perform poorly when the number of dimensions used to create the embedding is high.

\subsection{Music Genre Recommendation} 
Node embeddings are often used for extracting features of nodes for downstream predictive tasks. In order to investigate this, we use social networks of Deezer users collected from European countries. We predict the genres (out of $84$) of music liked by people. 
%Each graph was embedded with the standard parameter settings listed at the beginning of Section \ref{sec:experiments} and the parameters being tuned are set such that the clustering gives the highest modularity. 
Following the embedding, we used logistic regression with $\ell_2$ regularization to predict each of the labels and $90\%$ of the nodes were randomly selected for training. We evaluated the performance of the remaining users. Numbers reported in Table \ref{fig:pred_perfor} are $F_1$ scores calculated from $10$ experimental repetitions. $\textit{GEMSEC}_2$ significantly outperforms the other methods on all three countries' datasets. The performance advantage varies between $3.03\%$ and $4.95\%$. We also see that \textit{Smooth GEMSEC}$_2$ has lower accuracy, but it is able to outperform \textit{DeepWalk, LINE, Node2Vec, Walklets, ComE, M-NMF} and \textit{DANMF} on all datasets. %It should be noted that the good performance of \textit{Walklets} is misleading as that model has 5 times more free parameters than other models listed here. It is more expressive which might explain the advantage in terms of $F_1$ scores over other baselines. In addition, we have evidence that introducing the clustering cost and smoothness regularization does not affect performance on the downstream predictive task adversely. On the contrary, it can increase predictive accuracy significantly.

\begin{figure}[!h]
	\centering
	\begin{tikzpicture}[scale=0.28,transform shape]
	\tikzset{font={\fontsize{22pt}{12}\selectfont}}
	\begin{axis}[
	legend columns=3,
	width=1.0\textwidth,
	height=1.0\textwidth,
	grid=major,
	grid style={dashed, gray!40},
	scaled ticks=false,
	inner axis line style={-stealth},
	ymin=-3,
	ymax=13,
	ytick={-3,-1,...,7,9,11,13},
	xtick={7,9,11,13,15},
	legend style = {at={(0.5,-0.1)},anchor=north, column sep = 10pt,legend columns = -1},
	xlabel=$\log_{2}$ Vertex number,
	ylabel=$\log_{2}$ Optimization runtime (s),
	enlargelimits=0.2, 
	]
	
	\addplot[mark=*,opacity=0.8,mark options={black,fill=red},mark size=3pt]
	coordinates {
		
		(6,-2.993)
		(7,-1.861)
		(8,-0.959)
		(9,0.034)
		(10,1.157)
		(11,2.083)
		(12,2.968)
		(13,4.035)
		(14,5.131)
		(15,6.359)

	};\addlegendentry{DeepWalk}
	
	\addplot[mark=square*,opacity=0.8,mark options={black,fill=green},mark size=3pt]
	coordinates {
		
		(6,-2.585)
		(7,-1.649)
		(8,-0.68)
		(9,0.359)
		(10,1.34)
		(11,2.343)
		(12,3.375)
		(13,4.344)
		(14,5.4)
		(15,6.595)

	};\addlegendentry{Smooth DeepWalk}
	
	\addplot[mark=triangle*,opacity=0.8,mark options={black,fill=yellow},mark size=5pt]
	coordinates {
		
		(6,-2.852)
		(7,-1.851)
		(8,-0.874)
		(9,0.108)
		(10,1.265)
		(11,2.112)
		(12,3.087)
		(13,4.098)
		(14,5.193)
		(15,6.425)

	};\addlegendentry{GEMSEC}
	\addplot[mark=diamond*,opacity=0.8,mark options={black,fill=blue},mark size=5pt]
	coordinates {
		
		(6,-2.535)
		(7,-1.49)
		(8,-0.566)
		(9,0.508)
		(10,1.648)
		(11,2.395)
		(12,3.397)
		(13,4.437)
		(14,5.589)
		(15,6.679)
		
	};\addlegendentry{Smooth GEMSEC}
	
	\addplot[mark=*,opacity=0.8,mark options={black,fill=gray},mark size=3pt]
	coordinates {
		
		(6, -0.878452)
		(7,  -0.33484)
		(8,  0.721695)
		(9,  1.898808)
		(10,  3.184307)
		(11,  4.807687)
		(12,  6.832890)
		(13,  8.707386)
		(14,  10.790870)
		(15,12.759547)
	};\addlegendentry{M-NMF}
	
	\addplot[mark=diamond*,opacity=0.8,mark options={black,fill=gray},mark size=5pt]
	coordinates {
		
		(6,0.62685)
		(7,1.23372)
		(8,1.9982)
		(9,2.98341)
		(10,4.04508)
		(11,4.9834)
		(12,5.9846)	
		(13,6.9596)
		(14,7.93726)
		(15,8.9717)
		
	};\addlegendentry{ComE}
	
	\addplot[opacity=0.8,dashed,thick,mark options={black,fill=gray},mark size=5pt]
	coordinates {
		
		(6,-3.5)
		(7,-2.5)
		(8,-1.6)
		(9,-0.5)
		(10,0.5)
		(11,1.5)
		(12,2.5)
		(13,3.5)
		(14,4.5)
		(15,5.5)
		
	};
	
	\addplot[opacity=0.8,dashed,thick,mark options={black,fill=gray},mark size=5pt]
	coordinates {
		
		(6,-2)
		(7,-1)
		(8,-0)
		(9,1)
		(10,2)
		(11,3)
		(12,4)
		(13,5)
		(14,6)
		(15,7)
		
	};

	\end{axis}
	\end{tikzpicture}
	\caption[Sensitivity of GEMSEC optimization runtime to graph size.]{Sensitivity of optimization runtime to graph size measured by seconds. The dashed lines are linear references.}\label{fig:performance}
\end{figure}
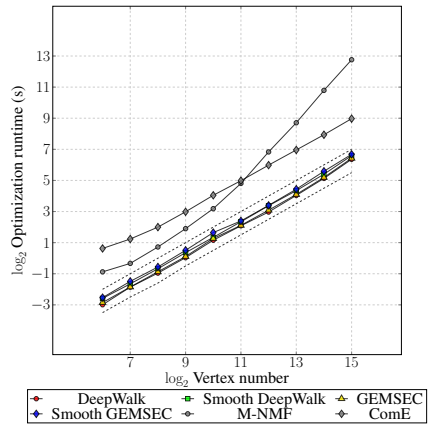

\subsection{Scalability and computational efficiency} To create graphs of various sizes, we used the Erdos-Renyi model and with an average degree of $20$. Figure \ref{fig:performance} shows the log of mean runtime against the log of the number of nodes. Most importantly, we can conclude that doubling the size of the graph doubles the time needed for optimizing \textit{GEMSEC}, thus the growth is linear. We also observe that embedding algorithms that incorporate clustering have a higher cost, and regularization also produces a higher cost, but similar growth.

\begin{table*}[htbp!]
		\caption[Multi-label node classification performance of the embedding extracted features on the Deezer genre likes datasets.]{Multi-label node classification performance of the embedding extracted features on the Deezer genre likes datasets. Performance is measured by average $F_1$ score values. Models were trained on 90\% of the data and evaluated on the remaining 10\%. Errors in the parentheses correspond to two standard deviations. \textit{GEMSEC} models consistently have good performance.}
		\label{fig:pred_perfor}

	\centering{\footnotesize
		\begin{tabular}{l ccc ccc ccc}
			& \multicolumn{3}{c}{\textbf{\textsc{Croatia}}} & \multicolumn{3}{c}{\textsc{\textbf{Hungary}}} & \multicolumn{3}{c}{\textsc{\textbf{Romania}}} \\ 		\cmidrule(l){2-4}\cmidrule(l){5-7}\cmidrule(l){8-10}
			& \textbf{Micro}  & \textbf{Macro}  & \textbf{Weighted}  & \textbf{Micro}  & \textbf{Macro}  & \textbf{Weighted}  & \textbf{Micro}  & \textbf{Macro}  & \textbf{Weighted}   \\\hline
			\textbf{Overlap Factorization}&$\underset{(\pm 0.017)}{0.319}$&$\underset{(\pm 0.002)}{0.026}$&$\underset{(\pm 0.010)}{0.208}$&$\underset{(\pm 0.007)}{0.361}$&$\underset{(\pm 0.001)}{0.029}$&$\underset{(\pm 0.006)}{0.227}$&$\underset{(\pm 0.025)}{0.275}$&$\underset{(\pm 0.003)}{0.020}$&$\underset{(\pm 0.017)}{0.167}$\\	[0.95em]	
			\textbf{DeepWalk}            & $\underset{(\pm0.006)}{0.321}$&	$\underset{(\pm0.002)}{0.026}$&	$\underset{(\pm0.004)}{0.207}$&	$\underset{(\pm0.004)}{0.361}$	&$\underset{(\pm0.002)}{0.029}$&	$\underset{(\pm0.002)}{0.228}$&	$\underset{(\pm0.008)}{0.307}$&	$\underset{(\pm0.002)}{0.023}$	&$\underset{(\pm0.006)}{0.186}$\\[0.95em]
			\textbf{LINE}&$\underset{(\pm0.013)}{0.331}$&$\underset{(\pm0.002)}{0.028}$&$\underset{(\pm0.010)}{0.212}$&$\underset{(\pm0.007)}{0.374}$&$\underset{(\pm0.002)}{0.033}$&$\underset{(\pm0.005)}{0.250}$&$\underset{(\pm0.007)}{0.332}$&$\underset{(\pm0.002)}{0.028}$&$\underset{(\pm0.006)}{0.212}$\\[0.95em]
			\textbf{Node2Vec} &$\underset{(\pm0.012)}{0.348}$&$\underset{(\pm0.003)}{0.032}$&$\underset{(\pm0.010)}{0.235}$&$\underset{(\pm0.008)}{0.393}$&$\underset{(\pm0.002)}{0.037}$&$\underset{(\pm0.011)}{0.267}$&$\underset{(\pm0.008)}{0.346}$&$\underset{(\pm0.002)}{0.031}$&$\underset{(\pm 0.008)}{0.229}$\\[0.95em]
			\textbf{Walklets}&$\underset{(\pm 0.013)}{0.363}$&$\underset{(\pm 0.003)}{0.043}$&$\underset{(\pm 0.012)}{0.270}$&$\underset{(\pm 0.007)}{0.397}$&$\underset{(\pm 0.001)}{0.051}$&$\underset{(\pm 0.006)}{0.307}$&$\underset{(\pm 0.011)}{0.361}$&$\underset{(\pm 0.005)}{\textbf{0.050}}$&$\underset{(\pm 0.012)}{0.281}$\\[0.95em]
			\textbf{ComE}&$\underset{(\pm 0.012)}{0.326}$&$\underset{(\pm 0.002)}{0.028}$&$\underset{(\pm 0.009)}{0.217}$&$\underset{(\pm 0.010)}{0.363}$&$\underset{(\pm 0.001)}{0.033}$&$\underset{(\pm 0.007)}{0.246}$&$\underset{(\pm 0.008)}{0.323}$&$\underset{(\pm 0.001)}{0.028}$&$\underset{(\pm 0.006)}{0.212}$\\[0.95em]
			\textbf{M-NMF}&$\underset{(\pm 0.005)}{0.336}$&$\underset{(\pm 0.001)}{0.028}$&$\underset{(\pm 0.003)}{0.217}$&$\underset{(\pm 0.015)}{0.369}$&$\underset{(\pm 0.002)}{0.032}$&$\underset{(\pm 0.011)}{0.239}$&$\underset{(\pm 0.016)}{0.330}$&$\underset{(\pm 0.002)}{0.028}$&$\underset{(\pm 0.013)}{0.209}$\\[0.95em]
			\textbf{DANMF}&$\underset{(\pm 0.007)}{0.340}$&$\underset{(\pm 0.002)}{0.027}$&$\underset{(\pm 0.002)}{0.210}$&$\underset{(\pm 0.011)}{0.365}$&$\underset{(\pm 0.002)}{0.031}$&$\underset{(\pm 0.008)}{0.242}$&$\underset{(\pm 0.009)}{0.335}$&$\underset{(\pm 0.002)}{0.029}$&$\underset{(\pm 0.012)}{0.210}$\\[0.95em]\hline
			\textbf{Smooth DeepWalk}            &$\underset{(\pm0.006)}{0.329}$&	$\underset{(\pm0.002)}{0.028}$&	$\underset{(\pm0.006)}{0.215}$&	$\underset{(\pm0.006)}{0.375}$	&$\underset{(\pm0.002)}{0.032}$&	$\underset{(\pm0.004)}{0.244}$&	$\underset{(\pm0.008)}{0.321}$&	$\underset{(\pm0.002)}{0.026}$	&$\underset{(\pm0.006)}{0.204}$\\[0.95em]
			\textbf{GEMSEC}            &	$\underset{(\pm0.006)}{0.328}$&	$\underset{(\pm0.002)}{0.027}$&	$\underset{(\pm0.004)}{0.212}$&	$\underset{(\pm0.004)}{0.377}$	&$\underset{(\pm0.002)}{0.032}$&	$\underset{(\pm0.004)}{0.244}$&	$\underset{(\pm0.008)}{0.332}$&	$\underset{(\pm0.002)}{0.028}$&	$\underset{(\pm0.006)}{0.213}$\\[0.95em]
			\textbf{Smooth GEMSEC}            &	$\underset{(\pm0.006)}{0.333}$&	$\underset{(\pm0.002)}{0.028}$&	$\underset{(\pm0.004)}{0.215}$&	$\underset{(\pm0.006)}{0.379}$	&$\underset{(\pm0.002)}{0.034}$&	$\underset{(\pm0.004)}{0.250}$&	$\underset{(\pm0.008)}{0.334}$&	$\underset{(\pm0.002)}{0.029}$&	$\underset{(\pm0.006)}{0.215}$\\[0.95em] \hline
			\textbf{GEMSEC}$_2$            &	$\underset{(\pm0.007)}{\textbf{0.381}}$&	$\underset{(\pm0.003)}{\textbf{0.046}}$&	$\underset{(\pm0.005)}{\textbf{0.287}}$&	$\underset{(\pm0.005)}{0.407}$	&$\underset{(\pm0.003)}{0.050}$&	$\underset{(\pm0.007)}{0.310}$&	$\underset{(\pm0.009)}{\textbf{0.378}}$&	$\underset{(\pm0.003)}{0.049}$&	$\underset{(\pm0.007)}{\textbf{0.289}}$\\[0.95em]
			\textbf{Smooth GEMSEC}$_2$             &	$\underset{(\pm0.005)}{0.373}$&	$\underset{(\pm0.002)}{0.044}$&	$\underset{(\pm0.006)}{0.276}$&	$\underset{(\pm0.004)}{\textbf{0.409}}$	&$\underset{(\pm0.002)}{\textbf{0.053}}$&	$\underset{(\pm0.006)}{\textbf{0.314}}$&	$\underset{(\pm0.008)}{0.376}$&	$\underset{(\pm0.003)}{0.049}$&	$\underset{(\pm0.007)}{0.287}$\\[0.95em] \hline 	
			
		\end{tabular}}	
\end{table*}

\section{Conclusions}\label{sec:conclusion}
We described \textit{GEMSEC} -- a novel algorithm that learns a node embedding and a clustering of nodes jointly. It extends existing embedding modes. We showed that smoothness regularization is used to incorporate social network properties and produce natural embedding and clustering. We presented new social datasets, and experimentally, our methods outperform a number of strong community aware node embedding baselines.

\section{Acknowledgements}
Benedek Rozemberczki and Ryan Davies were supported by the Centre for Doctoral Training in Data Science, funded by EPSRC (grant EP/L016427/1).

%\FloatBarrier 
\bibliographystyle{IEEEtran}
\bibliography{main}

\begin{thebibliography}{10}
\providecommand{\url}[1]{#1}
\csname url@samestyle\endcsname
\providecommand{\newblock}{\relax}
\providecommand{\bibinfo}[2]{#2}
\providecommand{\BIBentrySTDinterwordspacing}{\spaceskip=0pt\relax}
\providecommand{\BIBentryALTinterwordstretchfactor}{4}
\providecommand{\BIBentryALTinterwordspacing}{\spaceskip=\fontdimen2\font plus
\BIBentryALTinterwordstretchfactor\fontdimen3\font minus
  \fontdimen4\font\relax}
\providecommand{\BIBforeignlanguage}[2]{{%
\expandafter\ifx\csname l@#1\endcsname\relax
\typeout{** WARNING: IEEEtran.bst: No hyphenation pattern has been}%
\typeout{** loaded for the language `#1'. Using the pattern for}%
\typeout{** the default language instead.}%
\else
\language=\csname l@#1\endcsname
\fi
#2}}
\providecommand{\BIBdecl}{\relax}
\BIBdecl

\bibitem{van2012robust}
T.~Van~Laarhoven and E.~Marchiori, ``Robust community detection methods with
  resolution parameter for complex detection in protein protein interaction
  networks,'' \emph{Pattern Recognition in Bioinformatics}, pp. 1--13, 2012.

\bibitem{backstrom2006group}
L.~Backstrom, D.~Huttenlocher, J.~Kleinberg, and X.~Lan, ``Group formation in
  large social networks: Membership, growth, and evolution,'' in
  \emph{Proceedings of the 12th ACM SIGKDD international conference on
  Knowledge discovery and data mining}.\hskip 1em plus 0.5em minus 0.4em\relax
  ACM, 2006, pp. 44--54.

\bibitem{papadopoulos2012community}
S.~Papadopoulos, Y.~Kompatsiaris, A.~Vakali, and P.~Spyridonos, ``Community
  detection in social media,'' \emph{Data Mining and Knowledge Discovery},
  vol.~24, no.~3, pp. 515--554, 2012.

\bibitem{leskovec2014mining}
J.~Leskovec, A.~Rajaraman, and J.~D. Ullman, \emph{Mining of Massive
  Datasets}.\hskip 1em plus 0.5em minus 0.4em\relax Cambridge University Press,
  2014.

\bibitem{walktrap}
P.~Pascal and M.~Latapy, \emph{In International Symposium on Computer and
  Information Sciences}.\hskip 1em plus 0.5em minus 0.4em\relax Springer Berlin
  Heidelberg, 2005, ch. Computing Communities in Large Networks Using Random
  Walks., pp. 284--293.

\bibitem{gregory2010finding}
S.~Gregory, ``Finding overlapping communities in networks by label
  propagation,'' \emph{New Journal of Physics}, vol.~12, no.~10, p. 103018,
  2010.

\bibitem{goyal2017graph}
P.~Goyal and E.~Ferrara, ``Graph embedding techniques, applications, and
  performance: A survey,'' \emph{arXiv preprint arXiv:1705.02801}, 2017.

\bibitem{deepwalk}
B.~Perozzi, R.~Al-Rfou, and S.~Skiena, ``Deepwalk: Online learning of social
  representations.'' in \emph{Proceedings of the 20th ACM SIGKDD international
  conference on Knowledge discovery and data mining.}, 2014.

\bibitem{grover2016node2vec}
A.~Grover and J.~Leskovec, ``Node2vec: Scalable feature learning for
  networks,'' in \emph{Proceedings of the 22nd ACM SIGKDD International
  Conference on Knowledge Discovery and Data Mining}, 2016, pp. 855--864.

\bibitem{zachary1977information}
W.~W. Zachary, ``An information flow model for conflict and fission in small
  groups,'' \emph{Journal of anthropological research}, vol.~33, no.~4, pp.
  452--473, 1977.

\bibitem{mikolov_1}
T.~Mikolov, K.~Chen, G.~Corrado, and J.~Dean, ``Efficient estimation of word
  representations in vector space,'' 2013.

\bibitem{adampaper}
J.~B. Diederik P.~Kingma, ``Adam: A method for stochastic optimization,'' in
  \emph{Proceedings of the 3rd International Conference on Learning
  Representations (ICLR)}, 2015.

\bibitem{cavallari2017learning}
S.~Cavallari, V.~W. Zheng, H.~Cai, K.~C.-C. Chang, and E.~Cambria, ``Learning
  community embedding with community detection and node embedding on graphs,''
  in \emph{Proceedings of the 2017 ACM on Conference on Information and
  Knowledge Management}, 2017, pp. 377--386.

\bibitem{wang2017community}
X.~Wang, P.~Cui, J.~Wang, J.~Pei, W.~Zhu, and S.~Yang, ``Community preserving
  network embedding.'' in \emph{AAAI}, 2017, pp. 203--209.

\bibitem{ye2018deep}
F.~Ye, C.~Chen, and Z.~Zheng, ``Deep autoencoder-like nonnegative matrix
  factorization for community detection,'' in \emph{Proceedings of the 27th ACM
  International Conference on Information and Knowledge Management}.\hskip 1em
  plus 0.5em minus 0.4em\relax ACM, 2018, pp. 1393--1402.

\bibitem{tang2015line}
J.~Tang, M.~Qu, M.~Wang, M.~Zhang, J.~Yan, and Q.~Mei, ``Line: Large-scale
  information network embedding,'' in \emph{Proceedings of the 24th
  International Conference on World Wide Web}, 2015, pp. 1067--1077.

\bibitem{perozzidontwalk}
B.~Perozzi, V.~Kulkarni, H.~Chen, and S.~Skiena, ``Don't walk, skip!: Online
  learning of multi-scale network embeddings,'' in \emph{Proceedings of the
  2017 IEEE/ACM International Conference on Advances in Social Networks
  Analysis and Mining 2017}, 2017, pp. 258--265.

\bibitem{fakcharoenphol2004tight}
J.~Fakcharoenphol, S.~Rao, and K.~Talwar, ``A tight bound on approximating
  arbitrary metrics by tree metrics,'' \emph{Journal of Computer and System
  Sciences}, vol.~69, no.~3, pp. 485--497, 2004.

\bibitem{matouvsek2002lectures}
J.~Matou{\v{s}}ek, \emph{Lectures on discrete geometry}.\hskip 1em plus 0.5em
  minus 0.4em\relax Springer, 2002, vol. 108.

\bibitem{yu2011spherical}
X.~Yu, X.~Ban, W.~Zeng, R.~Sarkar, X.~Gu, and J.~Gao, ``Spherical
  representation and polyhedron routing for load balancing in wireless sensor
  networks,'' in \emph{2011 Proceedings IEEE INFOCOM}.\hskip 1em plus 0.5em
  minus 0.4em\relax IEEE, 2011, pp. 621--625.

\bibitem{huang2014bounded}
K.~Huang, C.-C. Ni, R.~Sarkar, J.~Gao, and J.~S. Mitchell, ``Bounded stretch
  geographic homotopic routing in sensor networks,'' in \emph{IEEE INFOCOM
  2014-IEEE Conference on Computer Communications}.\hskip 1em plus 0.5em minus
  0.4em\relax IEEE, 2014, pp. 979--987.

\bibitem{sarkar2011low}
R.~Sarkar, ``Low distortion delaunay embedding of trees in hyperbolic plane,''
  in \emph{International Symposium on Graph Drawing}.\hskip 1em plus 0.5em
  minus 0.4em\relax Springer, 2011, pp. 355--366.

\bibitem{de2018representation}
C.~De~Sa, A.~Gu, C.~R{\'e}, and F.~Sala, ``Representation tradeoffs for
  hyperbolic embeddings,'' \emph{Proceedings of machine learning research},
  vol.~80, p. 4460, 2018.

\bibitem{zeng2010resilient}
W.~Zeng, R.~Sarkar, F.~Luo, X.~Gu, and J.~Gao, ``Resilient routing for sensor
  networks using hyperbolic embedding of universal covering space,'' in
  \emph{2010 Proceedings IEEE INFOCOM}.\hskip 1em plus 0.5em minus 0.4em\relax
  IEEE, 2010, pp. 1--9.

\bibitem{rozemberczki2018fast}
B.~Rozemberczki and R.~Sarkar, ``Fast sequence-based embedding with diffusion
  graphs,'' in \emph{International Workshop on Complex Networks}.\hskip 1em
  plus 0.5em minus 0.4em\relax Springer, 2018, pp. 99--107.

\bibitem{qiu2018network}
J.~Qiu, Y.~Dong, H.~Ma, J.~Li, K.~Wang, and J.~Tang, ``Network embedding as
  matrix factorization: Unifying deepwalk, line, pte, and node2vec,'' in
  \emph{Proceedings of the Eleventh ACM International Conference on Web Search
  and Data Mining}.\hskip 1em plus 0.5em minus 0.4em\relax ACM, 2018, pp.
  459--467.

\bibitem{gutmann2010noise}
M.~Gutmann and A.~Hyvarinen, ``Noise-contrastive estimation: A new estimation
  principle for unnormalized statistical models,'' in \emph{Proceedings of the
  Thirteenth International Conference on Artificial Intelligence and
  Statistics}, 2010, pp. 297--304.

\bibitem{onnela2007structure}
J.-P. Onnela, J.~Saram{\"a}ki, J.~Hyv{\"o}nen, G.~Szab{\'o}, D.~Lazer,
  K.~Kaski, J.~Kert{\'e}sz, and A.-L. Barab{\'a}si, ``Structure and tie
  strengths in mobile communication networks,'' \emph{Proceedings of the
  national academy of sciences}, vol. 104, no.~18, pp. 7332--7336, 2007.

\bibitem{ahmed2013distributed}
A.~Ahmed, N.~Shervashidze, S.~Narayanamurthy, V.~Josifovski, and A.~J. Smola,
  ``Distributed large-scale natural graph factorization,'' in \emph{Proceedings
  of the 22nd international conference on World Wide Web}, 2013, pp. 37--48.

\end{thebibliography}

\end{document}